\documentclass[reprint,amsmath,amssymb,aps,]{revtex4-2}

\usepackage{caption}
\usepackage[utf8]{inputenc}
\usepackage{graphicx}
\usepackage{xcolor}
\usepackage{placeins}

\begin{document}

    \title
    {The Butterfly Defect Effect in Cylindrical Shell Buckling: \\ Non-localized Interactions of Localized Imperfections}

    \author{Uba K. Ubamanyu}%
    \affiliation{%
        Flexible Structures Laboratory, Institute of Mechanical Engineering,\\
        \'{E}cole Polytechnique F\'{e}d\'{e}rale de Lausanne (EPFL),\\
        1015 Lausanne, Switzerland\\
    }%
    
    \author{Antoine Blond}%
    \affiliation{%
        Flexible Structures Laboratory, Institute of Mechanical Engineering,\\
        \'{E}cole Polytechnique F\'{e}d\'{e}rale de Lausanne (EPFL),\\
        1015 Lausanne, Switzerland\\
    }%
    
    \author{John W. Hutchinson}%
    \affiliation{%
        School of Engineering and Applied Sciences,\\
        Harvard University,\\
        02138 Cambridge, MA, USA\\
    }%
    
    \author{Pedro M. Reis}%
     \email{pedro.reis@epfl.ch}
    \affiliation{%
        Flexible Structures Laboratory, Institute of Mechanical Engineering,\\
        \'{E}cole Polytechnique F\'{e}d\'{e}rale de Lausanne (EPFL),\\
        1015 Lausanne, Switzerland\\
        }%
    

\begin{abstract}
    We investigate the buckling of axially compressed cylindrical shells containing localized defects, focusing on interactions among multiple defects and between them and the shell edges. Using high-fidelity finite-element simulations, we first characterize the parameter space of single-dimple sensitivity, defect-edge coupling, and pairwise defect-defect interactions. 
    Building on this deterministic baseline, we run stochastic simulations of shells with randomly distributed imperfections whose amplitudes are sampled from a log-normal distribution. 
    Post-buckling deformations form non-axisymmetric, butterfly-shaped patterns whose `\textit{wings}' reach far across the shell surface. This spatial extent induces unavoidable interactions with neighboring defects and the clamped edges, so that buckling is not necessarily initiated by the deepest defect. A larger defect population raises the likelihood of an extreme defect, producing a statistical size effect: as the mean number of defects grows, the mean knockdown factor decreases asymptotically and its variability decays exponentially. The butterfly defect effect qualitatively explains the scatter in historical experimental data by correlating the knockdown factor with the cylinder length-to-thickness ratio ($H/t$) rather than the radius-to-thickness ratio ($R/t$) alone, thereby establishing $H$ as a key parameter for stability alongside $R$ and $t$. Nonetheless, the knockdown factors obtained here remain well above those reported in historical experiments, indicating that the localized Gaussian dimple, though nearly the worst-case imperfection for spherical shells, is not so for cylinders.
\end{abstract}

\maketitle 

\section{Introduction}\label{sec_intro}
    
The structural integrity of thin-walled cylindrical shells is paramount in aerospace and civil engineering, as reflected in the extensive classical literature on shell stability, amassed over the past century. Detailed accounts of shell buckling theories and classical results can be found in several monographs~\cite{von_karman_buckling_1941, timoshenko_theory_1963, brush_buckling_1975, bushnell_computerized_1985, elishakoff_resolution_2014}. Yet, despite this extensive body of work, contemporary design guidelines and codes for thin-shell structures remain heavily reliant on empirical lower-bound data collected over the past half-century~\cite{seide_development_1960, weingarten_buckling_1968, weingarten_buckling_1969,  hilburger_buckling_2020}. It is now well established that the severe discrepancy between theoretical predictions and experimental buckling loads stems from initial geometric imperfections, a sensitivity first rationalized by Koiter's seminal theory of elastic stability~\cite{koiter_over_1945}.
    
Buckling strength is customarily quantified by the knockdown factor $\kappa$, defined as the ratio of the observed buckling load to the classical theoretical prediction for a corresponding perfect shell. Based on the classic linear eigenproblem of small-deflection theory~\cite{lorenz_achsensymmetrische_1908, timoshenko_einige_1910, southwell_general_1914}, the predicted axial buckling load $P_c$ and critical end-shortening $\Delta_c$ for a geometrically perfect, linear-elastic cylindrical shell are given by
    \begin{equation}
        P_c = \frac{2\pi E t^2}{\sqrt{3(1-\nu^2)}} \quad \text{and} \quad \Delta_c = \frac{t H }{R\sqrt{3(1 - \nu^2)}}
        \label{Eq_Pc_Deltac}
    \end{equation}
where $E$ is  Young's modulus, $\nu$ is Poisson's ratio, and $R$, $t$, and $H$ are the radius, thickness, and height of the shell, respectively.

Historically, imperfection sensitivity was often assessed using global patterns or combinations of eigenmodes spanning the entire shell surface~\cite{carlson_experimental_1966, von_karman_buckling_1941, koiter_effect_1963, arbocz_initial_1979}. Because such global patterns are overly conservative and unrealistic representations of manufacturing flaws, the focus has shifted toward localized geometric imperfections such as discrete dimples and bumps~\cite{wullschleger_numerical_2006, lee_geometric_2016, gerasimidis_establishing_2018, abbasi_comparing_2023}.
    Cylindrical shells buckle by first forming a single localized dimple that develops into an axially localized but circumferentially periodic mode~\cite{hunt_localized_1991,horak_cylinder_2006}. Under continued compression, homoclinic snaking generates additional rings of axially localized dimples~\cite{hunt_localized_1991, kreilos_fully_2017, groh_role_2019}.
    
Early experimental studies assumed cylinder height ($H$) exerted minimal influence, plotting $\kappa$ solely against the radius-to-thickness ratio ($R/t$)~\cite{weingarten_elastic_1965, seide_development_1960}. However, perfect cylindrical shells with identical $R/t$ but distinct values of $H$ exhibit different critical buckling loads~\cite{wullschleger_numerical_2006}. This geometric dependency contributes to the scatter frequently observed in empirical datasets and establishes $H$ as a necessary parameter for characterizing stability limits~\cite{wullschleger_numerical_2006, wagner_robust_2017, groh_role_2019}.
    
Batdorf~\cite{batdorf_simplified_1947} formalized this geometric coupling through the dimensionless parameter
    \begin{equation}
        Z = \frac{H^2}{R t} \sqrt{1 - \nu^2}.
        \label{eq:Batdorf}
    \end{equation}
    
Physically, $Z$ compares the cylinder height to the buckling wavelength. The theoretical half-wavelength of the classical axisymmetric buckling mode~\cite{hutchinson_imperfection_1967} is 
    \begin{equation}
        l_c = \pi[12(1-\nu^2)]^{-1/4}\sqrt{Rt}.
        \label{eq:lc}
    \end{equation}
Combining Eq.~\eqref{eq:Batdorf} and Eq.~\eqref{eq:lc} yields $Z \propto (H/l_c)^2$. Collapsing the three-dimensional geometric space into the single axis, $Z$ establishes an equivalence class among cylindrical shells, excluding limiting cases of exceptionally thick, short, or column-like structures. Having defined this unified geometric scaling for ideal cylinders, we now turn to how initial imperfections degrade this idealized stability.
    
In practice, manufactured shells exhibit a stochastic spatial distribution of discrete imperfections. Because buckling resistance depends on the specific amplitude, geometry, and placement of these defects, the knockdown factor $\kappa$ is a random variable rather than a deterministic property. Probabilistic models applied to \textit{spherical} shells show that defects separated beyond a wavelength-dependent critical threshold act independently~\cite{derveni_defect-defect_2023, derveni_probabilistic_2023, baizhikova_probabilistic_2026}. Consequently, for a sufficiently large number of non-interacting defects, structural strength becomes an extreme-value statistics problem governed by the deepest individual defect, resulting in a finite weakest-link behavior~\cite{baizhikova_uncovering_2024, baizhikova_probabilistic_2026, le_statistical_2024}. The knockdown factors converge to a three-parameter Weibull distribution. This weakest-link behavior yields a universal statistical size effect governed by the radius-to-thickness ratio $R/t$~\cite{baizhikova_probabilistic_2026}. As the surface area grows, the probability of encountering a severe dimple rises, reducing the mean buckling capacity.
    
Whether an equivalent statistical scaling dictates the buckling of axially compressed cylindrical shells remains an open question. Recent high-fidelity simulations demonstrated that the defect interaction depends strongly on both defect spacing and orientation, and proposed an elliptical non-interaction envelope oriented at 30$^\circ$ to the circumferential axis~\cite{liu_interaction_2026}. However, their parametric scope was primarily constrained to discrete pairs of identical dimples or strictly axially aligned non-identical defects. While this finding confirms that cylindrical imperfections can behave independently under specific geometric conditions, the extent to which spherical probabilistic models translate to cylindrical structures requires further quantification. This change in geometry, from spheres to cylinders, raises a central research question: Do cylindrical shells exhibit an analogous weakest-link behavior and statistical size effect, and if so, what geometric parameters govern the scaling?
    
Here, we present the results of a comprehensive finite-element (FEM) simulation campaign investigating the buckling of cylindrical shells with localized imperfections. We structure our investigation by progressively increasing the complexity of the defect landscape. First, we explore the parameter space of cylindrical shells with a single Gaussian defect to establish a deterministic baseline. Second, building on the recent work by Liu \textit{et al.}~\cite{liu_interaction_2026}, we systematically explore the defect-defect interaction parameter space by analyzing a detailed matrix of relative orientations, spacings, and amplitude ratios. This broader approach reveals the physical mechanism underlying the interaction: at buckling onset, each defect develops a non-axisymmetric, `\textit{butterfly}'-shaped deformation pattern whose wings reach far across the shell surface, so that neighboring defects couple through the overlap of these patterns. We refer to this non-localized coupling as the \textit{butterfly defect effect}. Finally, we simulate shells containing stochastic distributions of multiple defects. We use these artificially generated stochastic imperfections to quantify the statistical size effects, where `\textit{size}' refers to the number of localized imperfections on the shell surface.
    
The remainder of this paper is organized as follows. We outline the problem definition and geometric constraints (Sec.~\ref{sec_probdef}) and detail the FEM methodology (Sec.~\ref{sec_FEM}). We then characterize the imperfection sensitivity of cylindrical shells with a single localized dimple (Sec.~\ref{sec_single_defect}), investigate defect-defect interactions (Sec.~\ref{sec_DefectDefect}), present the stochastic simulations with multiple defects (Sec.~\ref{sec_multiDefects}), and conclude (Sec.~\ref{sec_conclusions}).

\section{Problem Definition}\label{sec_probdef}
    
\begin{figure}[h!]
        \centering
        \includegraphics[width=0.9\linewidth]{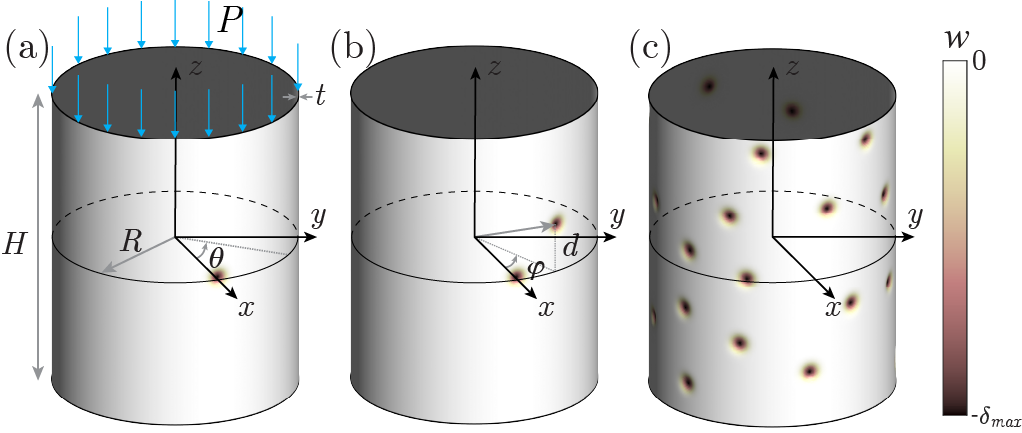}
        \caption{
        Cylindrical shells with localized Gaussian-dimple defects. 
        (a) Mid-surface radial deviation $w$ for a single centered defect $(\overline{\delta}=2, a=1, \lambda=1)$; color shows $w$ normalized by the maximum amplitude $\delta_{\mathrm{max}}$.
        (b) Two dimples separated by angular ($\varphi$) and axial ($d$) distances.
        (c) Stochastic multi-defect distribution.
        }
        \label{Fig_prob_def}
    \end{figure}

We consider a thin-walled cylindrical shell clamped at both of its upper and lower edges, and subjected to axial compression. The global geometry is set by the mid-surface radius $R$, thickness $t$, and height $H$; see Fig.~\ref{Fig_prob_def}(a). Two dimensionless parameters describe this geometry: the radius-to-thickness ratio $R/t$, and the Batdorf parameter $Z$ [Eq.~\eqref{eq:Batdorf}]. 

Geometric imperfections are introduced as a radial deviation, $w(\theta,z) = \sum_{i=1}^{N} w_i(\theta,z)$, from the nominal mid-surface of the perfect cylindrical shell, comprising $N$ localized Gaussian-dimple defects. 
Fig.~\ref{Fig_prob_def}(a-c) illustrates the initial geometry of imperfect shells with $N=1$, 2, and 30 defects, respectively; the latter illustrates the $N\gg 1$ case. Each defect, centered at coordinates $(\theta_i, z_i)$, is parameterized by the profile
    \begin{equation}
        w_i(\theta, z) = \overline{\delta}_i t \exp\left[-\left(\frac{2R(\theta-\theta_i)}{a_i\lambda_i l_c}\right)^{2} -\left(\frac{2(z-z_i)}{\lambda_i l_c}\right)^{2}\right],
        \label{Eq_Gaussian}
    \end{equation}
where $\overline{\delta}_i = \delta_i/t$ is the normalized amplitude and $l_c$ was given in Eq.~\eqref{eq:lc}. The $1/e^{4}$ ($\approx2\%$) contour defines a axial half-height $l_z=\lambda_i l_c$ and a circumferential half-width $l_x\approx a_i \lambda_i l_c$, so $\lambda_i$ sets the axial size relative to $l_c$, and $a_i=l_x/l_z$ is the aspect ratio. 
Our primary objective is to quantify how one or more localized geometric imperfections govern the buckling strength of cylindrical shells. Furthermore, we seek to assess how imperfection sensitivity translates from an isolated defect to a stochastic distribution of multiple defects. To address this, we structure our investigation around three distinct defect configurations.

First, we consider shells containing a single ($N=1$), localized Gaussian dimple [Fig.~\ref{Fig_prob_def}(a)] and quantify how the shell parameters ($R/t$, $Z$) and the defect parameters ($\overline{\delta}$, $\lambda$, $a$, and axial position) affect the buckling mode and the knockdown factor. This analysis identifies the parameter regions that yield consistent buckling modes, providing a basis for selecting parameters in the $N=2$ and $N \gg 1$ scenarios described next.
    
Second, we introduce a secondary defect to investigate defect-defect interactions [Fig.~\ref{Fig_prob_def}(b)]. The dimensionless angular distance $\varphi R/l_c$ and the axial distance $d/l_c$ parameterize the relative center-to-center separation between the two dimples. The central question for this configuration is whether a critical separation distance exists beyond which imperfections act independently, and how proximity raises or lowers the buckling strength.
    
Finally, we consider a more realistic scenario in which a shell contains a stochastic distribution of multiple imperfections ($N\gg 1$). The two key questions are (i) whether cylindrical shells exhibit weakest-link behavior, in which the most severe defect governs the buckling load, as observed in spherical shells~\cite{derveni_probabilistic_2023,baizhikova_uncovering_2024,baizhikova_probabilistic_2026}, and (ii) how the statistical size effect scales with the number of localized imperfections.
    
These three configurations are intrinsically linked, not merely ordered by increasing complexity. The single-defect results map out the parameter space for the two-defect study, and the resulting defect-pair interactions provide the geometric basis for interpreting the knockdown-factor statistics in the multi-defect stochastic regime. Resolving the localized snap-through instabilities common to all three configurations requires a FEM formulation tailored to unstable post-buckling paths, which we detail next.
    
\section{Methodology: FEM Simulations}\label{sec_FEM}

We conducted FEM simulations in the commercial package ABAQUS/Standard (2023) to systematically compute $\kappa$ across the $N=1$, $N=2$, and $N\gg1$ configurations introduced in the previous section. Following established modeling approaches for thin-shell buckling~\cite{lee_geometric_2016, jimenez_technical_2017, derveni_probabilistic_2023, ubamanyu_numerical_2025}, the shell was represented by its 3D mid-surface and discretized using four-node, reduced-integration shell elements (S4R). This discretization accurately captures buckling under axial compression while maintaining computational efficiency, thereby enabling thorough parameter exploration.

All simulations used a hyperelastic neo-Hookean model with $C_{10} = G/2=0.21\,$MPa, where $G$ is the initial shear modulus, and $D_1 = 0$ to enforce incompressibility. These values reproduce the response of VPS-32 silicone elastomer ($E = 1.26$~MPa, $\nu=0.5$) used in previous experimental studies~\cite{lee_geometric_2016, yan_buckling_2020,derveni_probabilistic_2023, ki_combined_2024, krida_vibration-assisted_2026}. Because imperfection sensitivity under elastic buckling is independent of the elastic modulus, the results carry over to other linear and nonlinear elastic materials.
    
The initial geometric imperfection, characterized by the Gaussian profile in Eq.~\eqref{Eq_Gaussian}, was introduced by radially adjusting the nodal coordinates of the otherwise perfect mid-surface mesh. To mimic classical experimental boundary conditions, the lower edge of the shell was completely clamped, while the upper edge was constrained against all degrees of freedom except axial translation. A uniform axial compressive force, equivalent to the classical buckling load $P_c$ [Eq.~\eqref{Eq_Pc_Deltac}], was applied to the upper edge. The analysis was performed using the arc-length Static/Riks method~\cite{riks_incremental_1979}, to capture the unstable snap-through post-buckling paths.
    
A detailed sensitivity analysis was performed to determine a suitable mesh density and solver configuration. Resolving the axisymmetric half-wavelength $l_c$ [Eq.~\eqref{eq:lc}] is critical, since it is the smallest geometric feature of interest. We define the discretization level as $m = l_c/d_m$, representing the number of finite elements spanning $l_c$, where $d_m$ is the approximate element size. For our nominal cylindrical geometry ($R = 40\,$mm, $t = 0.2\,$mm), convergence was achieved with an element size of $d_m \approx 0.64\,$mm, corresponding to $m = 8$.
    
Furthermore, the arc-length increment parameters ($\Delta s_{\text{ini}},\,\Delta s_{\text{min}},\,\Delta s_{\text{max}}$) of the Static/Riks solver were systematically evaluated. The initial and minimum increments were fixed at $\Delta s_{\text{ini}} = 0.01$ and $\Delta s_{\text{min}} = 10^{-10}$, respectively. Exploring the maximum admissible increment in the range $\Delta s_{\text{max}} \in [0.01, 0.1]$ revealed that $\Delta s_{\text{max}} = 0.02$ was sufficient for numerical stability throughout the solution path. The classical buckling load $P_c$ [Eq.~\eqref{Eq_Pc_Deltac}] was used as the nominal load throughout, so that the load proportionality factor equals $\kappa$ at the critical point.

Using this framework, we systematically explored a six-dimensional parameter space for shells with a single defect: the shell parameters $R/t \in [20,\,1500]$ and $Z \in [300,\,4000]$ (obtained by varying $R$, $H$, and $t$), the defect parameters $\overline{\delta} \in [0.1,\,2]$, $\lambda \in [0.5,\,3]$, $a \in [0.33,\,3]$, and their normalized axial position $z/(H/2) \in [0,\,1]$. The amplitude is examined throughout, the axial position in Sec.~\ref{sec_defectPlacement}, and the shape parameters $\lambda$ and $a$ in~\ref{App_DefectShapeSize}.
    
Finally, to investigate the stochastic distributions of multiple defects, we ran an ensemble of 200 deterministic FEM realizations for each geometric configuration, yielding more than 20{,}000 simulations in total. These were executed in parallel across multiple Dell Precision Tower 7920 workstations (dual Intel Xeon Gold 6248R processors, 24 cores / 48 threads each, 128~GB RAM), amounting to over 20,000 CPU-hours.

Following prior work~\cite{derveni_most_2025, derveni_probabilistic_2023, baizhikova_probabilistic_2026}, the localized imperfections were distributed across the cylindrical mid-surface using a random sequential adsorption algorithm, with defect amplitudes $\overline{\delta}_i$ drawn from a log-normal distribution of prescribed mean $\langle\overline{\delta}\rangle$ and standard deviation $\Delta\overline{\delta}$. For each dimple, the spatial coordinates (circumferential angle $\theta$ and axial position $z$) were sampled from a uniform distribution. To enforce a non-overlapping configuration, the geodesic distance between any two defect centers was constrained to exceed twice the characteristic half-width of the dimples, i.e. $2l_c$ for $a=1$, $\lambda=1$. Candidate locations violating this exclusion radius were rejected, and the process continued until either the prescribed number of defects, $N$, was seeded or the available surface area was exhausted. With the numerical framework and stochastic methodology established, we now characterize the imperfection sensitivity of a shell with a single localized dimple.

\section{Cylindrical Shells with a Single Localized Dimple}\label{sec_single_defect}

This section examines the local and global buckling modes of cylindrical shells with a single localized dimple, aiming to identify a geometry that fails by global buckling at its maximum load-carrying capacity, without prior localized instability, across a wide range of defect amplitudes. We first investigate how the Batdorf parameter ($Z$) governs imperfection sensitivity, then map the transition from local to global buckling. For a chosen geometry, we then characterize the spatial evolution of the distinctive `\textit{butterfly}' buckling pattern and assess how the proximity of defects to the clamped upper and lower edges affects the response.

\begin{figure}[!hbt]
        \centering   
        \includegraphics[width=0.9\linewidth]{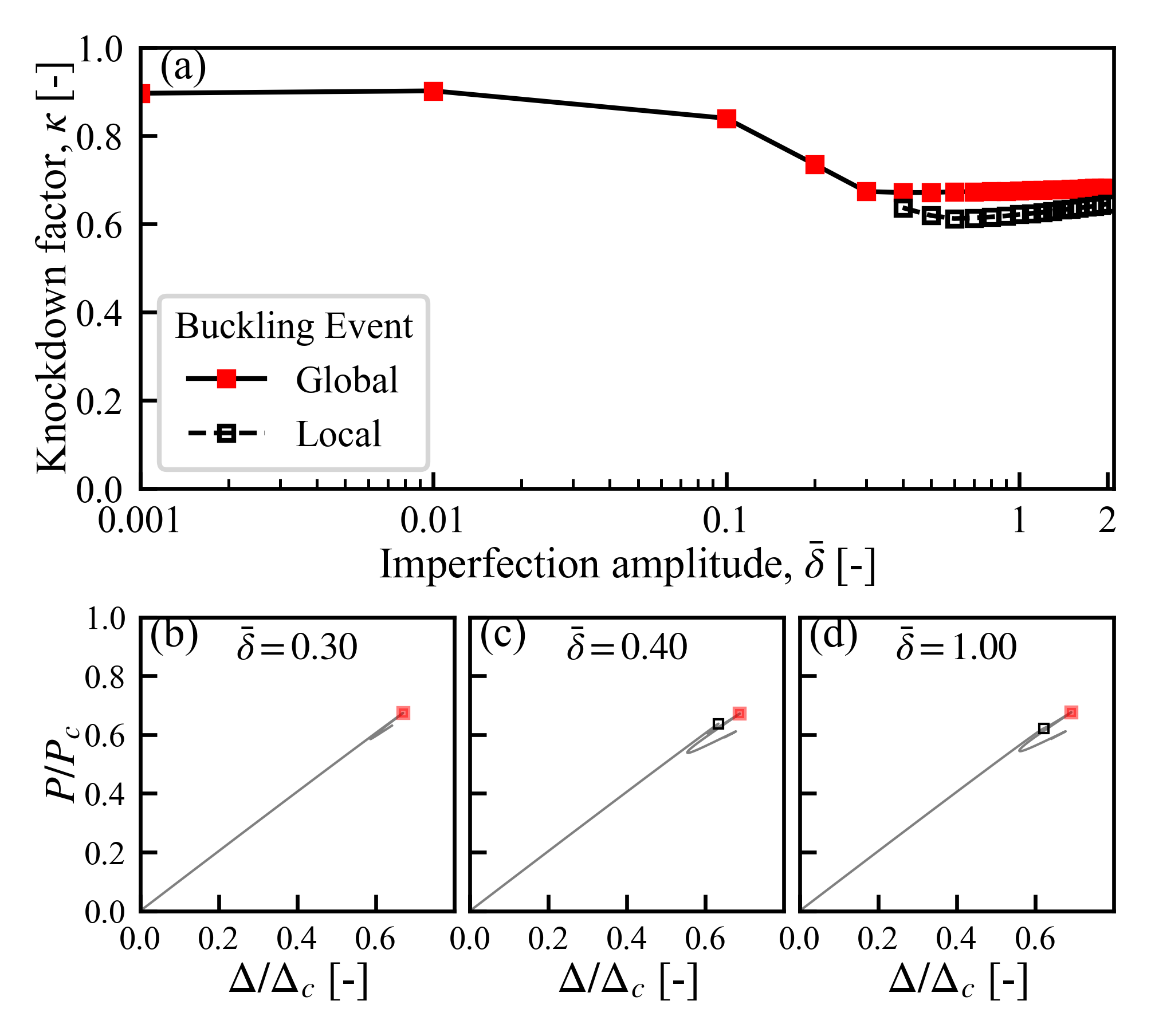}
        \caption{(a) Imperfection sensitivity curve, $\kappa(\overline{\delta})$, for a clamped cylindrical shell ($R/t = 160$, $H/R = 0.93$) with a localized Gaussian defect. Normalized load-displacement responses ($P/P_{c}$ versus normalized end-shortening $\Delta/\Delta_c$) for (b) $\overline{\delta} = 0.30$, (c) $\overline{\delta} = 0.40$, and (d) $\overline{\delta} = 1.00$. Markers denote local (open black squares) and global (filled red squares) buckling.}
        \label{typical_kdf-delta_Z300}
    \end{figure}

Figure~\ref{typical_kdf-delta_Z300}(a) presents a typical imperfection sensitivity curve ($\kappa$ versus $\overline{\delta}$) for a cylindrical shell with $R/t = 160$ and $H/R = 0.93$, containing a localized Gaussian defect of profile $(a,\, \lambda) = (1,\,1)$. In the near-perfect limit ($\overline{\delta} < 0.1$), $\kappa$ exhibits an upper-bound plateau at $\kappa \approx 0.9$, where buckling onset is dictated by the clamped edges rather than the localized defect. This mechanism parallels the edge-dominated buckling modes reported for near-perfect hemispherical shells~\cite{ubamanyu_numerical_2025}.
    
As the severity of the imperfection increases, the buckling mode transitions from immediate global collapse to a sequence initiated by localized instability. Two stability limits define these responses: local buckling, identified by a transient load drop or a sharp reduction in stiffness [open black squares in Fig.~\ref{typical_kdf-delta_Z300}(a)], and global buckling, which marks the ultimate structural collapse (filled red squares). The representative load-displacement insets in Fig.~\ref{typical_kdf-delta_Z300}(b-d) illustrate these distinct responses. Below the transition threshold ($\overline{\delta} < 0.4$), immediate global buckling occurs at the maximum load-carrying capacity, without prior localized instability [Fig.~\ref{typical_kdf-delta_Z300}(b)]. In contrast, shells with defect amplitudes exceeding this threshold ($\overline{\delta} > 0.4$) exhibit an initial, stable localized buckling mode prior to eventual global failure, consistent with earlier observations~\cite{haynie_validation_2012, gerasimidis_establishing_2018}. The local instability manifests as an initial load drop at moderate amplitudes [$\overline{\delta} = 0.40$, Fig.~\ref{typical_kdf-delta_Z300}(c)], and at higher amplitudes [$\overline{\delta} = 1.00$, Fig.~\ref{typical_kdf-delta_Z300}(d)] as a sharp decrease in stiffness rather than a load drop. The gap between the local and global buckling loads in Fig.~\ref{typical_kdf-delta_Z300}(a) represents the remaining load-carrying capacity following the initial local instability. Hereon, we consider only the first instability, whose load defines the critical knockdown factor $\kappa$.

\subsection{Influence of the $Z$ Parameter on Imperfection Sensitivity}

\begin{figure}[h!]
    \centering
    \includegraphics[width=0.9\linewidth]{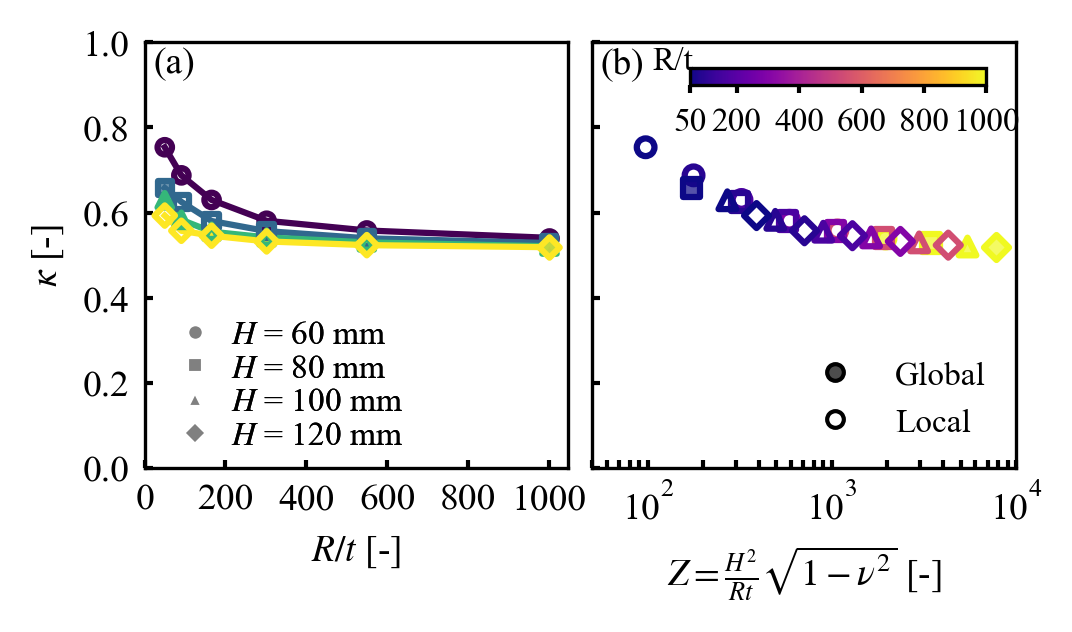}
    \caption{Critical knockdown factor $\kappa$ parameterized by cylinder geometry.
    (a) $\kappa$ versus radius-to-thickness ratio, $R/t$, for various cylinder heights, $H$, with a standardized defect $(\overline{\delta}, a, \lambda) = (2, 1, 1)$.
    (b) $\kappa$ versus the dimensionless Batdorf parameter, $Z$, plotted on a logarithmic axis.
    Marker shapes denote $H$, colormap indicates $R/t$, and fill denotes the buckling mode (solid for global, hollow for local).}
    \label{kappaEta_Z_collapse}
\end{figure}

For a standardized, centered Gaussian defect $(\overline{\delta}, a, \lambda) = (2, 1, 1)$, Fig.~\ref{kappaEta_Z_collapse}(a) displays $\kappa$ as a function of $R/t$ across varying heights and thicknesses. The complementary trends in aspect ratio and defect size, across a range of cylinder heights and defect amplitudes, are detailed in~\ref{App_DefectShapeSize} (Fig.~\ref{FigA_DefectSize_and_AspectRatio}); they justify the standardized profile $(a, \lambda) = (1, 1)$ used in our primary study to ensure a consistent buckling mode. The curves separate distinctly by cylinder height, highlighting the insufficiency of $R/t$ as the sole governing parameter. In contrast, when the same dataset is plotted against the Batdorf parameter $Z$ [Fig.~\ref{kappaEta_Z_collapse}(b)], the curves collapse onto a single master curve that decays monotonically toward a plateau.

\subsection{Mapping of the Local-to-Global Buckling Transition}

\begin{figure}[h!]
    \centering
    \includegraphics[width=0.9\linewidth]{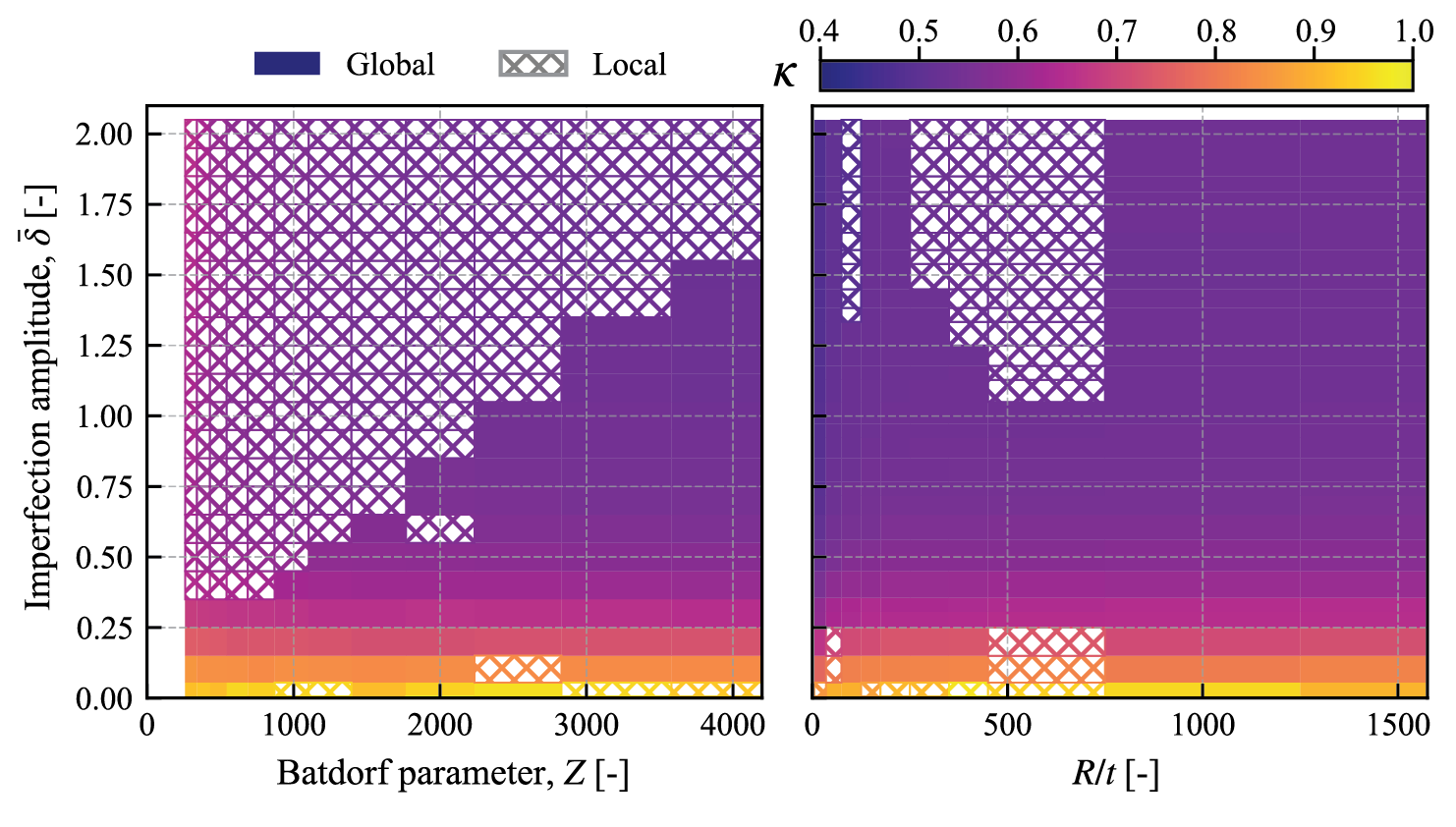}
    \caption{Phase diagram of $\kappa$ and the buckling mode for a cylinder with a localized Gaussian defect ($a = 1$, $\lambda = 1$): (a) the $Z$--$\overline{\delta}$ space at $R/t = 400$, and (b) the $R/t$--$\overline{\delta}$ space at $Z = 2771$. The colormap indicates $\kappa$. Overlays denote the critical buckling event: cross-hatched for localized dimple buckling and solid for immediate global collapse.}
    \label{fig_kdf_contour_Z_Rt_delta}
\end{figure}

Recognizing the governing role of the Batdorf parameter $Z$, we map $\kappa$ and the corresponding buckling mode across the $Z$--$\overline{\delta}$ parameter space. Figure~\ref{fig_kdf_contour_Z_Rt_delta}(a) shows this map for a standardized defect $(a, \lambda) = (1, 1)$ at constant $R/t = 400$, with the global mode as solid fill and the stable local mode as a cross-hatched overlay.

Consistent with classical imperfection sensitivity, $\kappa$ drops steeply and monotonically for small-amplitude defects ($\overline{\delta} \in [0, 0.5]$) across all tested geometries, before approaching a plateau. Notably, for cylinders with lower Batdorf parameters ($Z \lesssim 1500$), $\kappa$ shows a slight post-minimum recovery at extremely deep defects ($\overline{\delta} \in [1.0, 2.0]$).

The cross-hatched overlay in Fig.~\ref{fig_kdf_contour_Z_Rt_delta}(a) shows that the threshold defect amplitude required to trigger localized buckling increases monotonically with $Z$. For relatively short or thick cylinders (low $Z$), local buckling occurs at moderately deep defects ($\overline{\delta} \approx 0.5$). Conversely, highly slender or long cylinders (high $Z$) resist localized instability, sustaining global buckling even at severe defect amplitudes ($\overline{\delta} > 1.5$).

\vspace{-2mm}

To ensure that statistics from subsequent stochastic simulations reflect a uniform global-collapse mechanism, we selected a cylinder with $R = 40$~mm, $t = 0.2$~mm, $H = 160$~mm (i.e., $Z = 2771$, $R/t = 200$, $H/R = 4$) that exhibits global buckling across a wide range of defect amplitudes. The defect geometry is held fixed ($a = 1$, $\lambda = 1$) to isolate the effect of amplitude $\overline{\delta}$. To verify that this choice remains representative across varying $R/t$, Fig.~\ref{fig_kdf_contour_Z_Rt_delta}(b) shows the sensitivity of $\kappa$ to $R/t$ over the same range of $\overline{\delta}$. The response remains uniform at $R/t \approx 200$ and in the thin-shell regime ($R/t > 600$); however, local buckling occurs at large defect amplitudes in the intermediate range ($200 < R/t < 600$). The chosen geometry therefore provides a robust basis for isolating imperfection sensitivity from purely geometric scaling.

Having identified a global-collapse regime for the $Z = 2771$ cylinder, we fix the shell geometry to this configuration and analyze the spatial evolution of the localized buckling pattern as the defect amplitude increases.
     
\subsection{Evolution of Localized Butterfly Buckling Pattern} \label{sec_butterfly}

\begin{figure}[hbt!]
    \centering
    \includegraphics[width=0.85\linewidth]{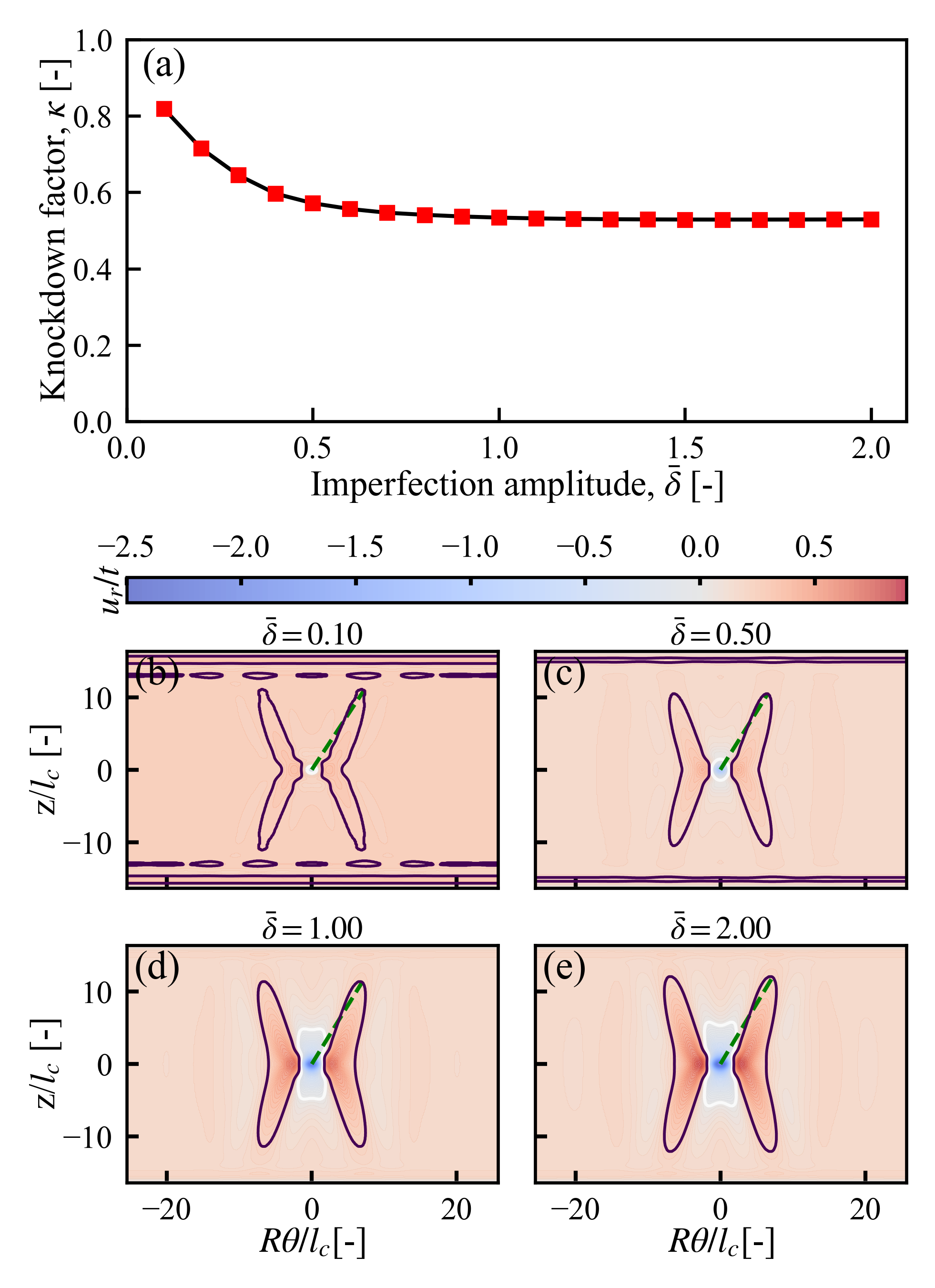}
    \caption{Buckling behavior of the selected cylindrical shell ($Z \approx 2771$).
    (a) Imperfection sensitivity curve illustrating $\kappa$ as a function of $\overline{\delta}$.
    (b--e) Contour maps of the unwrapped normalized radial displacement fields, $u_r/t$, at the onset of buckling for selected defect amplitudes $\overline{\delta}=\{0.10, 0.50, 1.00, 2.00\}$. The superimposed black and white lines show the primary buckling lobes (referred to as the \textit{wing of the butterfly}) and the zero-level contour. The dashed line highlights the wing's orientation relative to the defect center.}
    \label{fig_ChosenCylinder_Z2771}
\end{figure}

For the selected $Z \approx 2771$ cylinder, $\kappa$ decreases monotonically toward a plateau as $\overline{\delta}$ increases [Fig.~\ref{fig_ChosenCylinder_Z2771}(a)]. At the critical buckling load, the shell undergoes localized snap-through to form a stable, non-axisymmetric dimple. Figure~\ref{fig_ChosenCylinder_Z2771}(b-e) shows the radial displacement fields at buckling onset for $\overline{\delta}=\{0.1, 0.5, 1.0, 2.0\}$, respectively. The radial displacement field resembles a `\textit{butterfly}', with the inward-displacement region forming the body and two outward-displacement lobes extending diagonally as the wings. Superimposed zero-level contours (dashed white) and primary buckling lobes (solid black) in Fig.~\ref{fig_ChosenCylinder_Z2771}(b-e) delineate the spatial extent of the pattern, and a dashed radial line marks the wing-tip orientation.

These deformation patterns highlight the role of Gaussian curvature in buckling propagation. In spherical shells (positive Gaussian curvature), deformation remains strictly localized and axisymmetric~\cite{koiter_over_1945, audoly_localization_2020}; this localized response explains why spherical dimple imperfections interact only when their separation falls below the buckling half-wavelength $l_c$~\cite{derveni_defect-defect_2023, derveni_probabilistic_2023}. In contrast, cylindrical shells (zero Gaussian curvature) exhibit extensive spatial coupling: although earlier studies describe the initial buckled mode as ``localized''~\cite{groh_role_2019, kreilos_fully_2017}, the butterfly wings reach far toward the edges and decay only slowly with distance.

\begin{figure}[hbt!]
    \centering
    \includegraphics[width=0.85\linewidth]{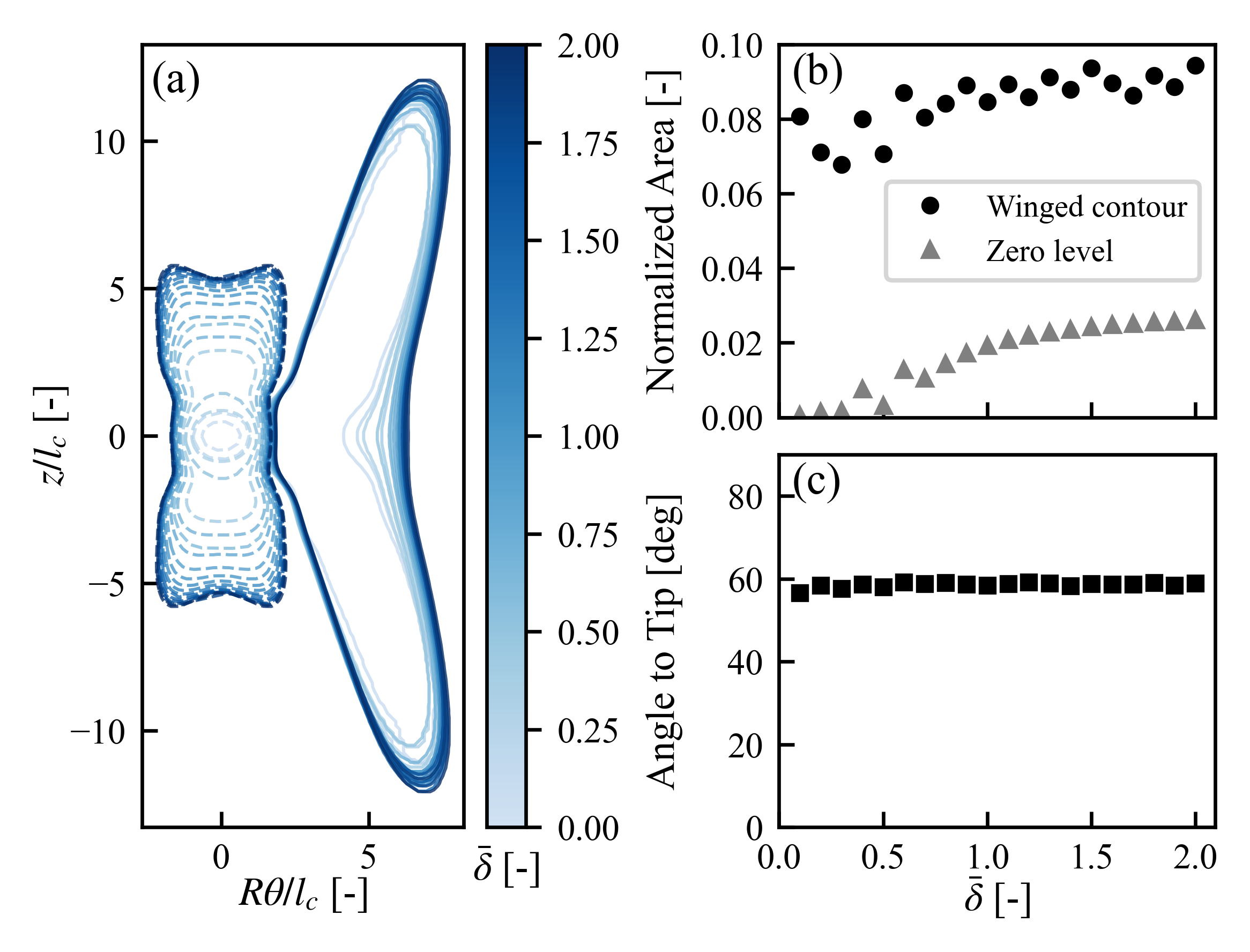}
    \caption{Evolution of the primary buckling-lobe characteristics with normalized defect amplitude $\overline{\delta}$.
    (a) Spatial evolution of the extracted wing and central-lobe contours on the normalized, unwrapped surface coordinates $(R\theta/l_c, z/l_c)$; solid and dashed lines show the primary ``wing'' and zero-level (central-lobe) contours, respectively, with the color gradient indicating $\overline{\delta}$. (b) Enclosed area of the ``wing'' and zero-level contours, normalized by the total shell surface area. (c) Orientation angle of the wing tip, measured from the circumferential axis.}
    \label{fig_butterfly_area_angle}
\end{figure}

Figure~\ref{fig_butterfly_area_angle}(a) maps the spatial evolution of the central lobe and one primary wing across $\overline{\delta} \in [0.1, 2.0]$. The central lobe traces the zero-displacement inflection contour, while the wing traces the primary outward displacement, isolated by subtracting the pre-buckling background deformation and thresholding the result (histogram-based, at 10\% above baseline) to capture the sharpest gradient.

Normalizing the enclosed contour areas by the total shell surface area ($2\pi R H$) confirms that the central lobe grows monotonically with amplitude, whereas the wing area stays roughly constant at around 8-10\% of the shell surface across the full amplitude range, already large at the smallest amplitudes [Fig.~\ref{fig_butterfly_area_angle}(b)]. By contrast, the wing tip stays oriented at approximately $60^{\circ}$ from the circumferential axis across all amplitudes [Fig.~\ref{fig_butterfly_area_angle}(c)]. Because this kinematic footprint spans a significant fraction of the cylinder, it is not strictly localized: its spatial reach unavoidably couples the dimple to the clamped edges and to any neighboring defects. We term this non-localized influence of a nominally localized imperfection the \textit{butterfly defect effect}; it underlies the defect-defect interactions (Sec.~\ref{sec_DefectDefect}) and the statistical size effect (Sec.~\ref{subsec:statistical_size}) characterized in the remainder of this study.

\subsection{Defect Placement and Edge Effects} \label{sec_defectPlacement}

\begin{figure}[h!]
    \centering
    \includegraphics[width=0.9\linewidth]{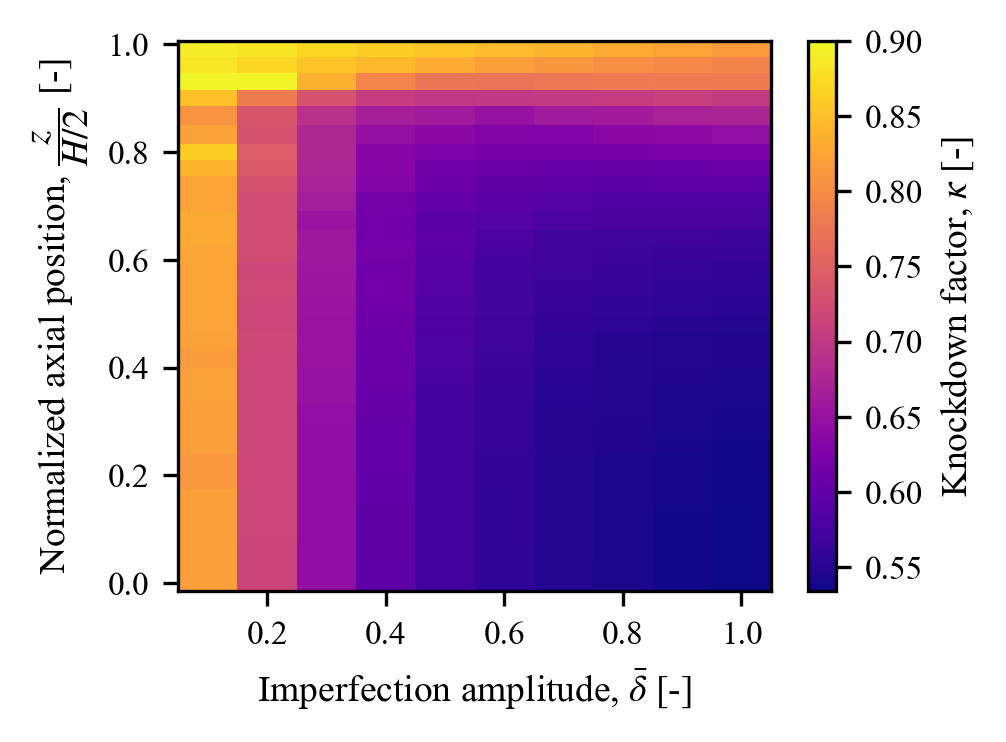}
    \caption{Knockdown factor $\kappa$ for a shell with a single Gaussian dimple, mapped across normalized defect amplitude $\overline{\delta}$ and axial position $z$ (scaled by half the cylinder height, $H/2$). The clamped edge makes $\kappa$ strongly dependent on placement, especially at higher amplitudes, where the edge effect reaches further inward from the edge.}
    \label{kdf_defectPlacement}
\end{figure}

The clamped edges locally stiffen the shell, preventing the butterfly pattern from fully developing. Consequently, the edges interact with the displacement field even when the imperfection is seeded far from them. Figure~\ref{kdf_defectPlacement} maps $\kappa$ across normalized axial positions and amplitudes $(z/(H/2), \overline{\delta}) \in [0, 1] \times [0.1, 1.0]$ for fixed shape parameters $(a, \lambda) = (1, 1)$. For shallow defects (small $\overline{\delta}$), $\kappa$ remains independent of axial position up to $z/(H/2) \approx 0.8$, indicating a highly localized edge effect. For deeper defects within the plateau region of the $\kappa(\overline{\delta})$ curve, proximity to the edge governs $\kappa$ over a substantially wider range. Defect size widens this reach as well: larger defect sizes $\lambda$ extend the edge effect further from the edge, as discussed in~\ref{App_BoundarySize} (Fig.~\ref{Fig_App_BoundarySize}).

This spatial sensitivity shows that a defect's severity depends on both its amplitude and its proximity to the clamped edge. To study defect-defect interactions systematically (Sec.~\ref{sec_DefectDefect}), we must isolate the localized instability from edge-stiffening effects. We therefore define an exclusion zone near the edges, ensuring that load-capacity reductions in multi-defect configurations arise solely from defect interactions.
    
\section{Defect-Defect Interaction}\label{sec_DefectDefect}

This section analyzes the interaction between two localized Gaussian defects [Fig.~\ref{Fig_prob_def}(b)], characterizing their spatial coupling and the constructive interference of their buckling modes. The findings inform the separation criterion adopted in the subsequent stochastic multi-defect simulations (Sec.~\ref{sec_multiDefects}).

\begin{figure}[h!]
    \centering
    \includegraphics[width=0.85\linewidth]{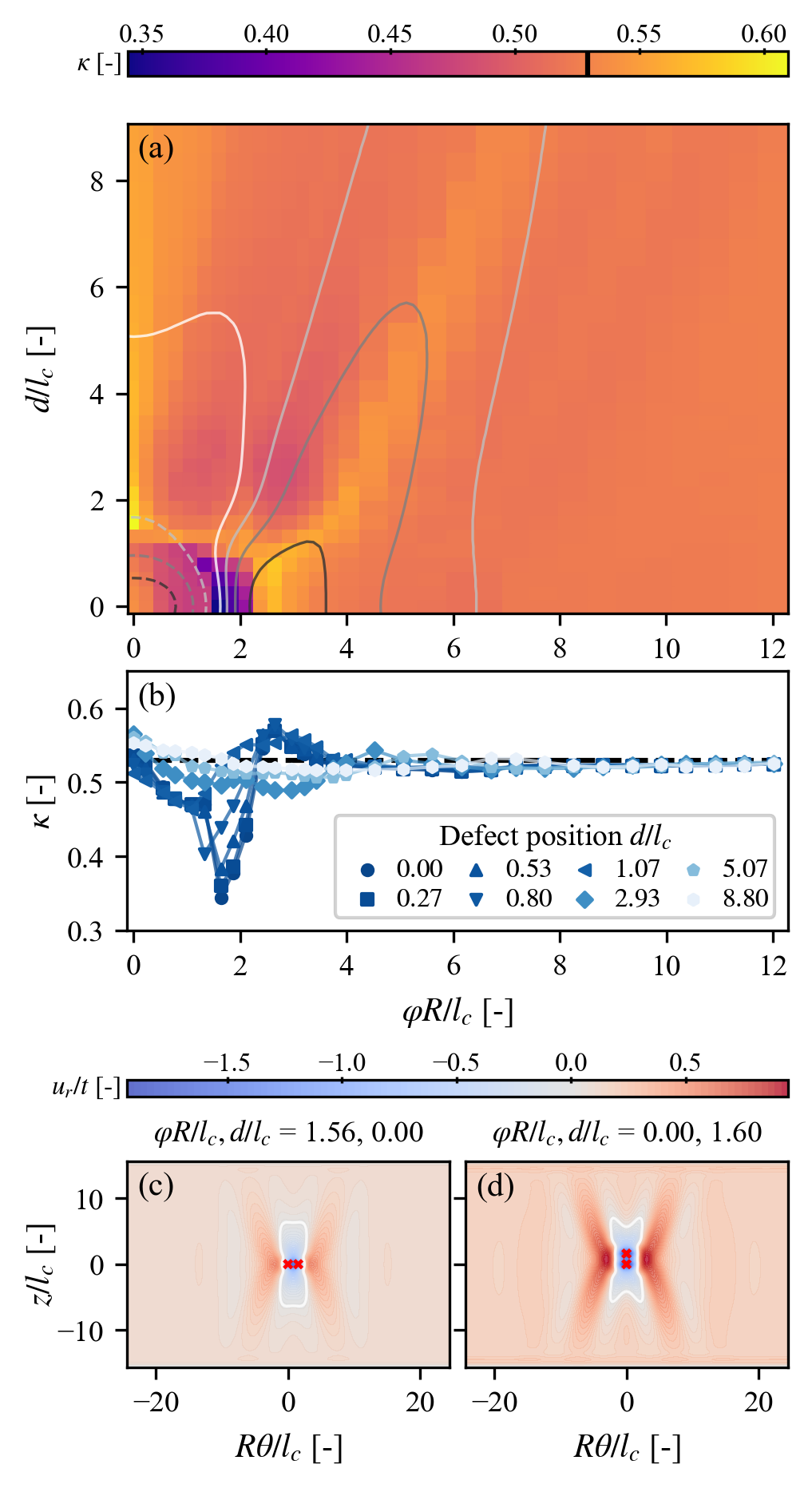}
    \caption{Interaction between two identical Gaussian defects ($\overline{\delta} = 2.0$). (a) Map of the knockdown factor $\kappa$ as a function of the normalized circumferential ($\varphi R/l_c$) and axial ($d/l_c$) separations. Solid and dashed lines mark the positive and negative contours of the single-defect radial deflection at buckling onset. (b) $\kappa$ versus circumferential separation for selected axial distances (see legend); the horizontal dashed line marks the single-defect value ($\kappa = 0.53$ at $\overline{\delta} = 2.0$). (c, d) Unwrapped normalized radial displacement fields, $u_r/t$, at buckling onset for two configurations marked by red crosses in (a): (c) purely circumferential separation ($\varphi R/l_c = 1.56$, $d/l_c = 0.00$) and (d) purely axial separation ($\varphi R/l_c = 0.00$, $d/l_c = 1.60$).}
    \label{Fig_DefectDefect}
\end{figure}

For two Gaussian dimples of fixed amplitude $\overline{\delta} = 2.0$, Fig.~\ref{Fig_DefectDefect}(a) maps $\kappa$ against the normalized circumferential ($\varphi R/l_c$) and axial ($d/l_c$) separations.
When the defects are close, a strong interaction zone forms, lowering the buckling load well below the single-defect value ($\kappa = 0.53$ at $\overline{\delta} = 2.0$). Rather than forming a circular region, the interaction zone extends diagonally outward. This contour closely mirrors the butterfly displacement pattern of a single defect (Sec.~\ref{sec_butterfly}), confirming that two defects interact most strongly when the secondary defect lies near the circumferential inflection of the first defect's displacement field. Figure~\ref{Fig_DefectDefect}(b) shows $\kappa$ as a function of circumferential separation for selected axial distances. For purely circumferential alignment ($d/l_c = 0$), $\kappa$ drops sharply, reaching a minimum of $\kappa \approx 0.35$ at $\varphi R/l_c \approx 1.6$, about a third below the isolated-defect baseline ($\kappa = 0.53$). As axial separation increases, the minimum-$\kappa$ location shifts diagonally in the $(\varphi R/l_c, d/l_c)$ space and the interaction weakens. For large axial separations, the interaction vanishes, consistent with the diagonal reach of the butterfly mode.

The radial displacement fields at buckling onset [Fig.~\ref{Fig_DefectDefect}(c,d)] reveal the physical origin of this orientation dependence. The severe strength reduction under circumferential alignment results from constructive interference of the localized buckling modes. Purely circumferential separation merges the individual butterfly patterns into a single, wider buckling mode [Fig.~\ref{Fig_DefectDefect}(c)] that acts as a larger, more severe geometric imperfection and buckles at a substantially lower load. Pure axial separation, by contrast, stacks the individual modes vertically [Fig.~\ref{Fig_DefectDefect}(d)]; because the primary lobes propagate diagonally, this alignment produces little constructive amplification and a milder strength reduction.

These results establish that relative defect positioning is as critical as defect amplitude in governing shell stability, motivating the stochastic multi-defect analysis in Sec.~\ref{sec_multiDefects}. We generalize this behavior to defects of unequal amplitudes in~\ref{App_UnequalDefects} (Fig.~\ref{Fig_App_UnequalDefects}), where the regions of strongest interaction align with the contours of the isolated single-defect radial deflection across all configurations.

Unlike spherical shells, which have a discrete non-interaction threshold, the extended diagonal reach of the cylindrical buckling mode precludes a strictly localized cutoff, a direct manifestation of the butterfly defect effect. The stochastic multi-defect simulations in Sec.~\ref{sec_multiDefects} therefore enforce a minimum spatial separation of $2\,l_c$. This geometric criterion keeps neighboring defects non-overlapping, even though mechanical interactions persist beyond that distance.
    
\section{Stochastic Simulations with Multiple Defects}
\label{sec_multiDefects}

Having studied the $N=1$ and $N=2$ configurations, we proceed to the stochastic FEM simulations of cylindrical shells containing multiple localized defects ($N \gg 1$). We first establish how the defect population scales with shell geometry, then test whether a weakest-link mechanism governs buckling, and finally quantify the statistical size effect and its dependence on the variance of defect amplitude. We close by reinterpreting the scatter in historical experimental data through the lens of our stochastic simulations.

\subsection{Scaling of the Defect Population}

Realistic shell structures contain many randomly distributed imperfections rather than isolated defects, motivating a shift from a deterministic to a probabilistic framework with $N \gg 1$ localized defects. In the subsequent simulations, we fix the Batdorf parameter at $Z = 2771$ and systematically alter the available surface area by varying the shell dimensions, as detailed in Table~\ref{Table_GeoSizeEffect}. The high $Z$ value ensures the buckling load remains insensitive to geometric scaling (Fig.~\ref{kappaEta_Z_collapse}), isolating the effect of defect population. Defects are distributed across the shell mid-surface using a random sequential adsorption algorithm, as detailed in Sec.~\ref{sec_FEM}, with amplitudes drawn from a log-normal distribution of prescribed mean $\langle\overline{\delta}\rangle$ and standard deviation $\Delta\overline{\delta}$.

\begin{table}[h!]
    \centering
       \caption{Geometric parameters of the cylindrical shells used in the stochastic multi-defect simulations. The Batdorf parameter is fixed at $Z=2771$ across all configurations. The table details the radius ($R$), height ($H$), and thickness ($t$), along with their corresponding dimensionless ratios ($R/t$ and $H/t$), and the $\overline{N}_{d, \text{saturated}}$ representing the mean defect count at the packing limit when defects are distributed across the full surface.}
    \begin{tabular}{ccccccccc}
        $R$ [mm] & $H$ [mm] & $t$ [mm] & $R/t$ [-] & $H/t$ [-] & $\overline{N}_{,\text{saturated}}$\\
        \hline \hline
        21.9       & 118.5      & 0.2        & 109.7  & 592.4  & 187.5             \\
        34.3       & 148.1      & 0.2        & 171.4  & 740.5  & 231.6             \\
        49.4       & 177.7      & 0.2        & 246.8  & 888.6  & 275.3             \\
        67.2       & 207.3      & 0.2        & 335.9  & 1036.7 & 319.1             \\
        87.7       & 237.0      & 0.2        & 438.7  & 1184.8 & 361.3             \\
        111.0      & 266.6      & 0.2        & 555.2  & 1332.9 & 403.7             \\
        137.1      & 296.2      & 0.2        & 685.5  & 1481.0 & 445.5             \\
        165.9      & 325.8      & 0.2        & 829.4  & 1629.1 & 485.7             \\
        197.4      & 355.4      & 0.2        & 987.1  & 1777.2 & 528.1             \\
        231.7      & 385.0      & 0.2        & 1158.4 & 1925.2 & 567.3            
    \end{tabular}
    \label{Table_GeoSizeEffect}
\end{table}

\begin{figure}[hbt!]
    \centering
    \includegraphics[width=0.8\linewidth]{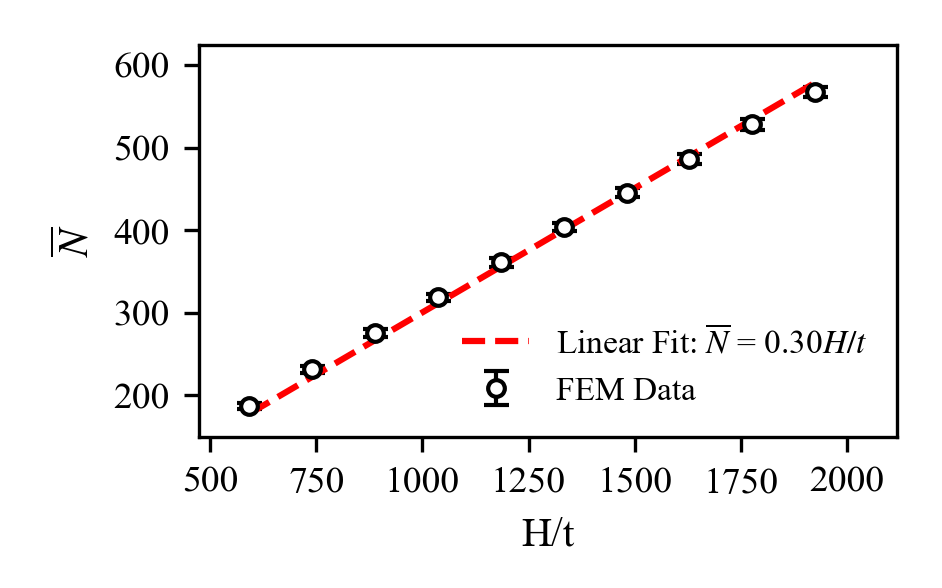}
    \caption{Mean number of seeded defects $\overline{N}$ at saturation versus the length-to-thickness ratio $H/t$. Black circles show FEM data; error bars give the standard deviation across realizations. The dashed red line is a linear fit ($\overline{N} \propto H/t$).}
    \label{Fig_Nd_vs_Ht}
\end{figure}

The number of defects $N$ that can be seeded is constrained by the available surface area and the non-overlapping criterion, which imposes a minimum center-to-center separation of $2\,l_c$ (for $a=1,\,\lambda=1$). We define the saturated state as the maximum defect count achievable before the surface area is exhausted. Since each defect occupies an area of order $l_c^2$ and the shell's surface area scales as $A \approx 2\pi RH$, the saturation count scales as
\begin{equation}
    N \sim \frac{\text{area of the shell}}{\text{area of the defect}} \sim \frac{2 \pi RH}{l_c^2} \sim \frac{H}{t}.
\end{equation}
Figure~\ref{Fig_Nd_vs_Ht} confirms this linear scaling, establishing $\overline{N}$ as a direct proxy for the available surface area. This relationship allows us to probe the statistical size effect either by scaling the shell geometry at saturation or by artificially capping the defect count below the packing limit.

\subsection{Testing the Weakest-Link Hypothesis}

\begin{figure}[h!]
    \centering
    \includegraphics[width=0.95\linewidth]{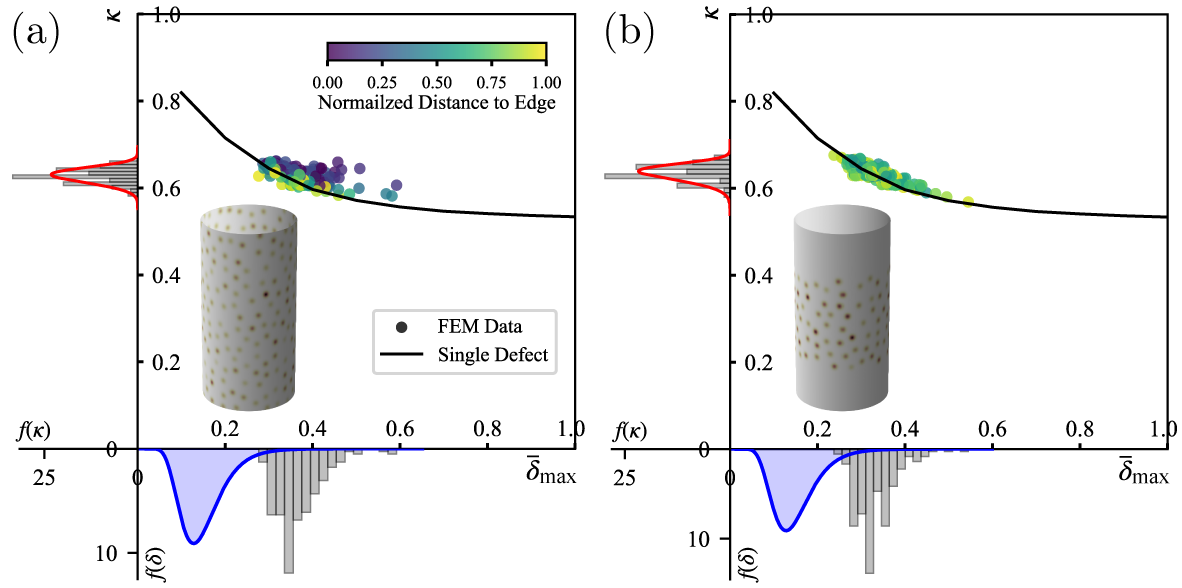}
    \caption{Knockdown factor $\kappa$ versus maximum seeded defect amplitude $\overline{\delta}_{\text{max}}$ for 200 stochastic realizations, compared against the single-defect sensitivity curve (solid black line), for defects seeded across (a) the full surface and (b) the mid-band ($z \in [-H/4, H/4]$). Marker colors indicate the normalized axial distance of the deepest defect from the nearest edge. The marginal plots show the probability density of $\overline{\delta}_{\text{max}}$ (bottom) and the Weibull-fitted distribution of $\kappa$ (left). Shell geometry: $(R,\,H,\,t) = (49.4,\,177.7,\, 0.2)$~mm ($Z = 2771$); defect distribution $(\langle \overline{\delta} \rangle,\, \Delta \overline{\delta}) = (0.15,\,0.05)$.}
    \label{fig:Compare_Full_vs_Half_shellarea}
\end{figure}

Having established the geometric scaling of the defect population, we investigate whether a weakest-link mechanism governs buckling; i.e., whether $\kappa$ of a multi-defect realization matches that of a single defect with the maximum seeded amplitude, $\overline{\delta}_{\text{max}}$. We consider a representative geometry ($R = 49.4$~mm, $H = 177.7$~mm, $t = 0.2$~mm, giving $Z = 2771$), with defect amplitudes drawn from a log-normal distribution ($\langle \overline{\delta} \rangle = 0.15$, $\Delta \overline{\delta} = 0.05$), and compare two configurations at the saturation limit: defects distributed across the full surface [Fig.~\ref{fig:Compare_Full_vs_Half_shellarea}(a)] and defects restricted to the mid-band [Fig.~\ref{fig:Compare_Full_vs_Half_shellarea}(b)].

For the fully populated configuration [Fig.~\ref{fig:Compare_Full_vs_Half_shellarea}(a)], the 200 stochastic realizations scatter broadly and deviate systematically from the single-defect sensitivity curve (solid black line). The markers, color-coded by the proximity of the deepest defect to the clamped edges, reveal the origin of this deviation: when the deepest defect sits near the edges, edge stiffening suppresses the expected localized instability and raises $\kappa$. In such cases, a shallower defect farther from the edge could govern failure instead, and the deepest defect alone does not dictate buckling.

To isolate the intrinsic imperfection sensitivity from edge effects, we restrict defect seeding to the mid-band of the cylinder ($z \in [-H/4, H/4]$); see Fig.~\ref{fig:Compare_Full_vs_Half_shellarea}(b). Eliminating edge interactions brings the data into close agreement with the single-defect curve, restoring localized failure. The remaining scatter reflects defect-defect interactions, as characterized in Sec.~\ref{sec_DefectDefect}.

The marginal plots of Fig.~\ref{fig:Compare_Full_vs_Half_shellarea} show the underlying statistical distributions: the bottom plots display the input log-normal distribution alongside the histogram of $\overline{\delta}_{\text{max}}$, while the left plots show the resulting (output) distribution of $\kappa$, fitted to a three-parameter Weibull. Notably, the two configurations yield qualitatively similar macroscopic Weibull statistics despite arising from fundamentally different mechanisms: edge-defect interactions in the fully populated shell and defect-defect interactions in the mid-band shell. This convergence shows that edge and defect-defect interactions are indistinguishable at the macroscopic statistical level, yet disentangling them remains necessary for predictive modeling.
    
\subsection{Statistical Size Effect}\label{subsec:statistical_size}

\begin{figure}[hbt!]
    \centering
    \includegraphics[width=0.8\linewidth]{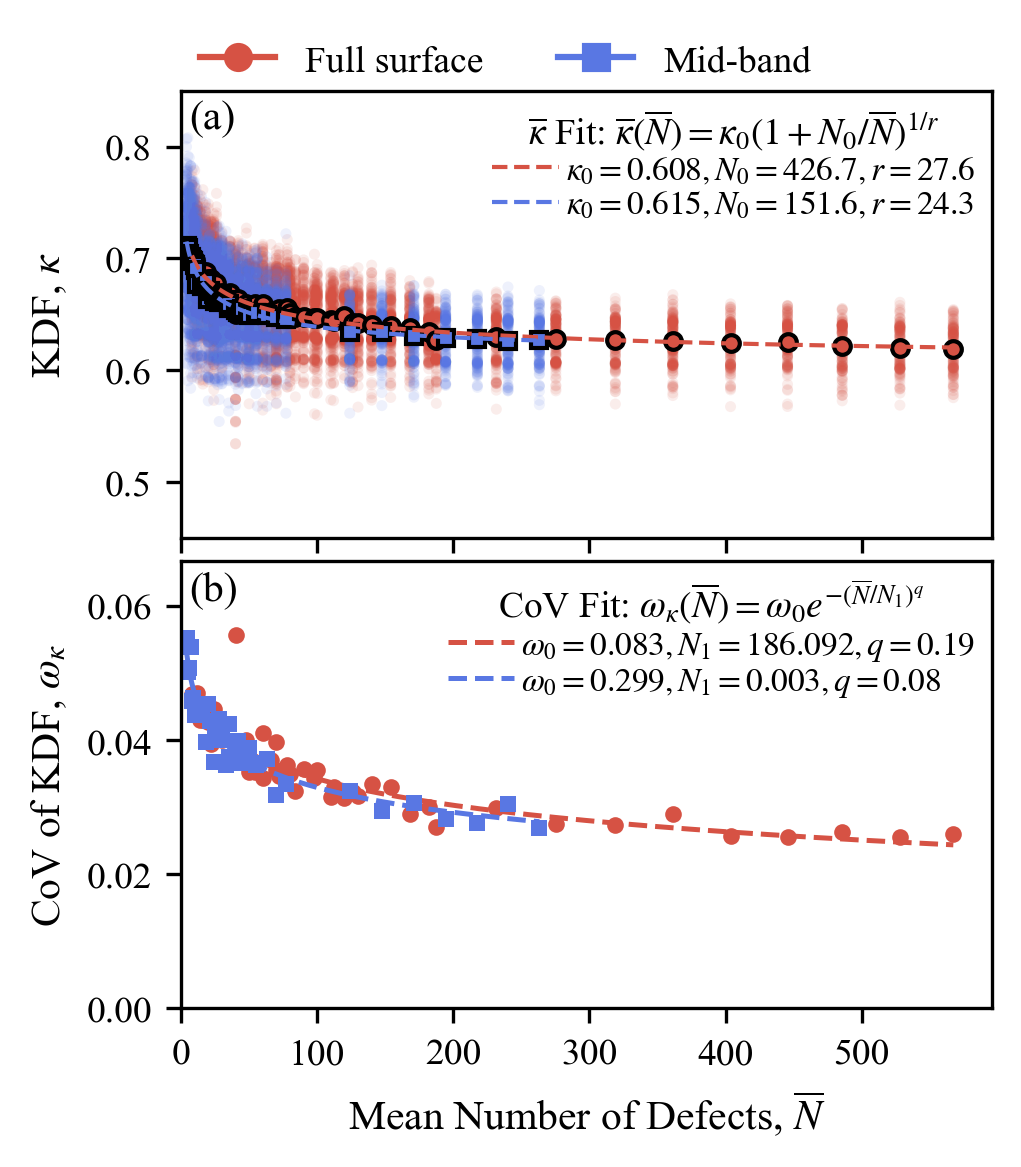}
    \caption{Statistical size effect on the buckling knockdown factor, comparing full-surface and mid-band defect distributions. (a) Mean knockdown factor $\overline{\kappa}$ versus mean defect count $\overline{N}$, overlaid on the scatter of 200 individual stochastic realizations; dashed lines show nonlinear regression fits. (b) Corresponding coefficient of variation $\omega_{\kappa}$ of the knockdown factor, fitted to an exponential decay model.}
    \label{fig_Size_Effect_Full_vsHalf}
\end{figure}

We now expand the analysis across various shell geometries ($R,\,t,\,H$), extracting the mean knockdown factor $\overline{\kappa}$ and the coefficient of variation $\omega_{\kappa}$ as functions of the mean defect count $\overline{N}$, for both configurations at their saturation limit and with artificially capped defect counts. Figure~\ref{fig_Size_Effect_Full_vsHalf}(a) plots $\overline{\kappa}$ for the fully populated (circular markers) and mid-band restricted (square markers) shells, with the 200 individual stochastic realizations overlaid as semi-transparent scatter.
As $\overline{N}$ increases, $\overline{\kappa}$ decreases and asymptotically approaches a lower bound. We quantify this decay using the scaling model from Ref.~\cite{baizhikova_probabilistic_2026}:
\begin{equation}
    \overline{\kappa}(\overline{N}) = \kappa_0\left(1+\frac{N_0}{\overline{N}}\right)^{1/r} 
    \label{mean_size_effect}
\end{equation}
where $\kappa_0$ is the asymptotic lower-bound plateau as $\overline{N} {\to} \infty$, $N_0$ is the characteristic defect count governing the onset of interaction-dominated failure, and $r$ controls the decay rate. Physically, $\kappa_0$ should approach the deterministic worst-case limit of the single-defect sensitivity curve $\kappa(\overline{\delta})$ as $\overline{N} {\to} \infty$. Regression yields $(\kappa_0,\,N_0,\,r) =(0.608,\,426.7,\,27.6)$ for the fully populated shell, and $(\kappa_0,\,N_0,\,r) = (0.615,\,151.6,\, 24.3)$ for the mid-band configuration. The substantially smaller $N_0$ in the mid-band case reflects the absence of edge stiffening: without edge shielding, a much smaller defect population suffices to trigger interaction-dominated failure. Nevertheless, both configurations converge to a nearly identical worst-case limit ($\kappa_0 \approx 0.61$), confirming this minimum is an intrinsic property of the shell geometry and the Gaussian-dimple imperfection rather than an artifact of boundary conditions.

The coefficient of variation $\omega_{\kappa}$, shown in Fig.~\ref{fig_Size_Effect_Full_vsHalf}(b), follows an exponential decay~\cite{baizhikova_probabilistic_2026}
\begin{equation}
    \omega_\kappa (\overline{N}) = \omega_0 \exp\left[-\left(\frac{\overline{N}}{N_1}\right)^{q}\right]
    \label{CoV_size_effect}
\end{equation} 
where $\omega_0$ is the coefficient of variation in the limit $\overline{N} {\to} 0$, $N_1$ is the characteristic population size governing the decay, and $q$ controls the decay's abruptness. Regression yields $(\omega_0,\,N_1,\,q) = (0.083,\,186,\,0.19)$ for the fully populated shell and $(\omega_0,\,N_1,\,q) = (0.299,\,0.003,\,0.08)$ for the mid-band configuration. The exponential decay of $\omega_{\kappa}$ confirms that, regardless of whether edge-defect or defect-defect interactions dominate locally, the structural response becomes increasingly deterministic as the defect population grows, with both configurations converging to indistinguishable macroscopic statistics.

\begin{figure}[hbt!]
    \centering
    \includegraphics[width=0.8\linewidth]{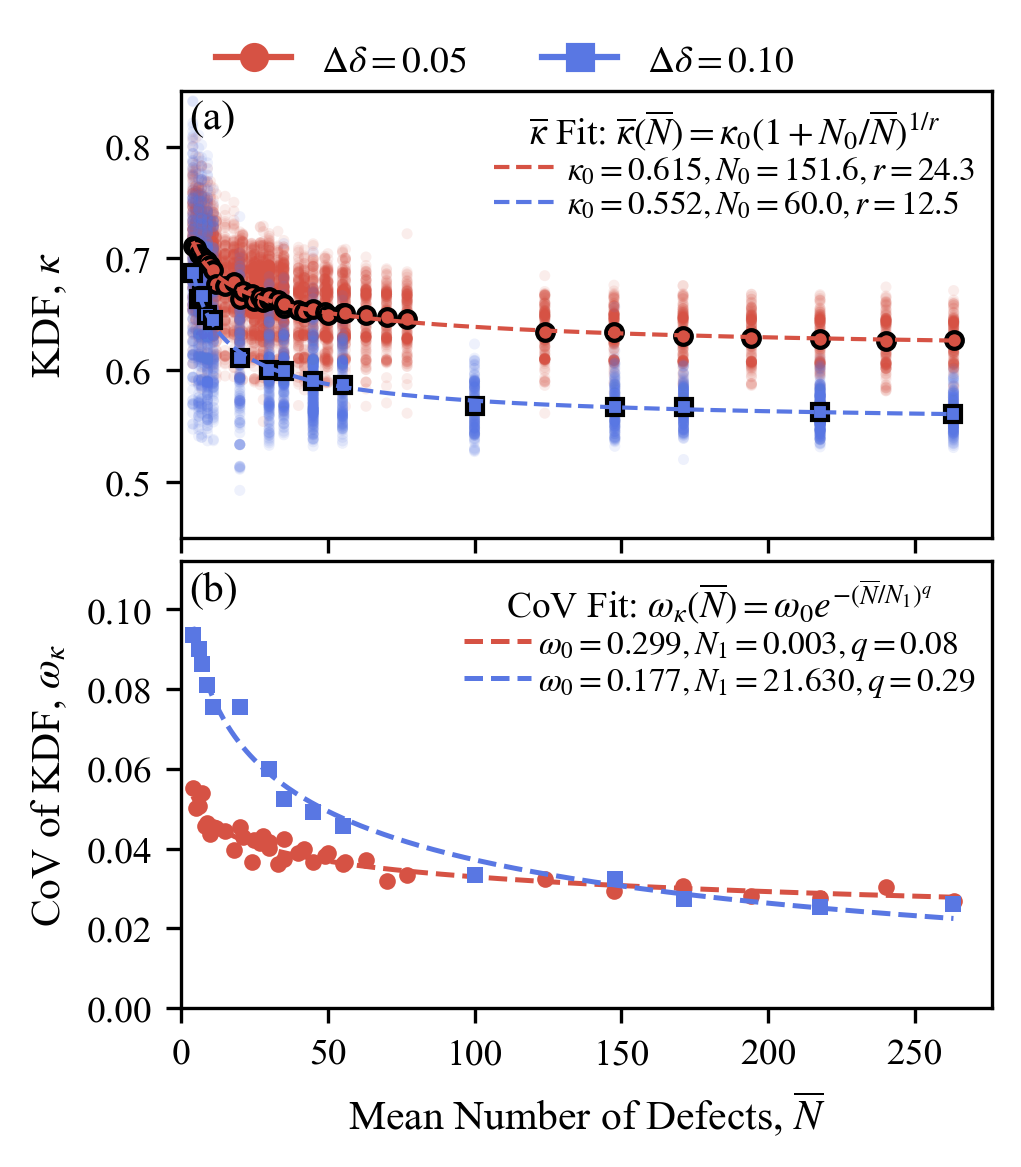}
    \caption{Statistical size effect on the buckling knockdown factor for mid-band defect distributions, comparing input log-normal standard deviations $\Delta\overline{\delta} = 0.05$ and $0.10$. (a) Mean knockdown factor $\overline{\kappa}$ versus mean defect count $\overline{N}$, overlaid on the scatter of 200 individual stochastic realizations; dashed lines show nonlinear regression fits. (b) Corresponding coefficient of variation $\omega_{\kappa}$, fitted to an exponential decay model in Eq.~\eqref{CoV_size_effect}.}
    \label{Fig_Size_Effect_Std}
\end{figure}

Having established that macroscopic statistical convergence is robust to the spatial configuration of defects, we now isolate the role of defect-amplitude variance in the size effect. Figure~\ref{Fig_Size_Effect_Std} compares size-effect statistics for shells populated from log-normal distributions with identical means ($\langle \overline{\delta} \rangle = 0.15$) but different standard deviations ($\Delta\overline{\delta} = 0.05$ and $0.10$), with defects restricted to the mid-band in both cases to eliminate edge stiffening effects.

The broader distribution ($\Delta\overline{\delta} = 0.10$) lowers the asymptotic worst-case limit, with regression yielding $(\kappa_0,\,N_0,\,r) = (0.552,\,60.0,\,12.5)$, compared to $(\kappa_0,\,N_0,\,r) = (0.615,\,151.6,\,24.3)$ for the narrower baseline ($\Delta\overline{\delta} = 0.05$). A wider amplitude distribution increases the probability of generating exceptionally deep defects within a single realization, lowering the asymptotic plateau and accelerating convergence toward it. This trend is equally reflected in $\omega_{\kappa}$ shown in Fig.~\ref{Fig_Size_Effect_Std}(b): fitting Eq.~\eqref{CoV_size_effect} yields $(\omega_0,\,N_1,\,q) = (0.177,\,21.63,\,0.29)$ for the broader distribution, compared to $(\omega_0,\,N_1,\,q) = (0.299,\,0.003,\,0.08)$ for the narrower baseline. The larger decay exponent ($q = 0.29$) confirms that amplitude variance governs the rate of statistical convergence: broader manufacturing scatter yields more extreme defect clusters, accelerating the system toward its deterministic limit. Together, these results show that the statistical variance of manufacturing flaws controls not only the mean load-carrying capacity but also the reliability scaling of the cylindrical shell.

These trends hold for dimples of fixed shape; coupling amplitude with defect size and shape is left to future work.
    
\subsection{Reinterpreting Scatter in Historical Experimental Data}
\label{sec_historical}

\begin{figure*}[hbt!]
    \centering
    \includegraphics[width=0.9\linewidth]{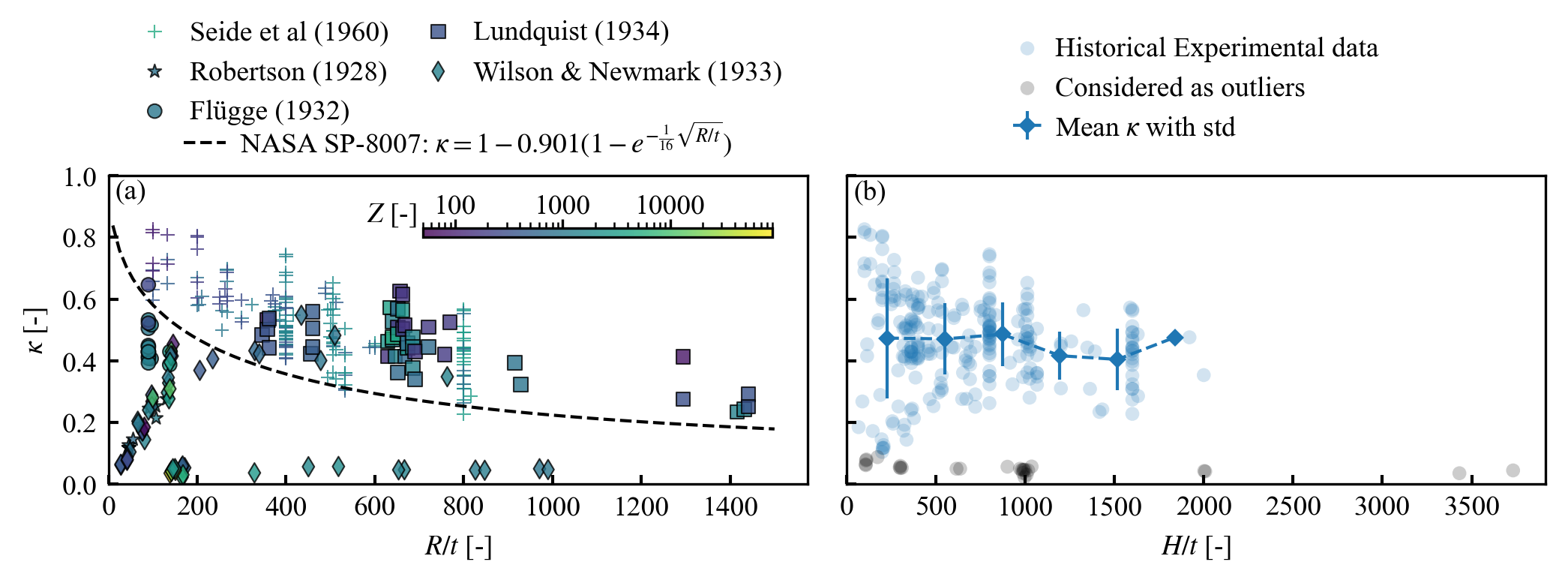}
    \caption{Historical experimental buckling data for cylindrical shells (see legend for references~\citet{seide_development_1960,robertson_strength_1928,flugge_stabilitat_1932,lundquist_strength_1934,wilson_strength_1933}). (a) Knockdown factor $\kappa$ versus $R/t$; color indicates the Batdorf parameter $Z$ on a logarithmic scale, revealing that shells with similar $R/t$ but distinct $Z$ scatter broadly. The dashed black line shows the NASA SP-8007 lower-bound design curve. (b) The same dataset replotted against $H/t$. Light blue markers show valid experimental data; gray markers identify outliers ($\kappa < 0.1$). The dashed line with diamond markers shows the binned mean knockdown factor and standard deviation.}
    \label{Fig_Historic_Data}
\end{figure*}

Figure~\ref{Fig_Historic_Data}(a) compiles historical experimental buckling data for cylindrical shells, plotted against $R/t$ following the convention of empirical design guidelines such as NASA SP-8007. Despite spanning decades of experiments, the dataset exhibits extensive scatter that $R/t$ alone fails to collapse.

We hypothesize that this scatter stems from two effects identified earlier. First, the deterministic buckling load is governed by the Batdorf parameter $Z$ rather than by $R/t$ alone (Sec.~\ref{sec_single_defect}), so shells that share an $R/t$ but differ in $Z$ carry different baseline knockdown factors. The $Z$ color-coding in Fig.~\ref{Fig_Historic_Data}(a) makes this explicit: at a given $R/t$, points separate systematically by $Z$, which is why a single-parameter curve cannot organize the data.
 
Second, the dispersion about this baseline is the statistical fingerprint of the butterfly defect effect: because the influence of each dimple is non-localized, the buckling load depends on the full defect arrangement rather than on the single deepest defect. Increasing $H/t$ enlarges the defect population ($\overline{N} \propto H/t$), so this interaction-driven scatter progressively averages out and $\omega_\kappa$ decays.
Replotting the historical data against $H/t$ [Fig.~\ref{Fig_Historic_Data}(b)] shows the corresponding signature: the experimental scatter about the binned mean narrows as $H/t$ grows, qualitatively consistent with the decay of $\omega_{\kappa}$ predicted by our simulations. Because the defect distributions of these experiments are unknown and the dataset spans a wide range of $Z$, whereas our $\omega_{\kappa}(\overline{N})$ fit was obtained at a single $Z$, a direct quantitative mapping is not possible. 

Even so, both mechanisms together provide a physics-based rationale for revisiting $R/t$-only lower bounds such as NASA SP-8007, and for treating cylinder length, through $Z$ and $H/t$, as an explicit stability parameter. Quantitative validation would require dedicated experimental campaigns that document both manufacturing imperfections and buckling loads, alongside a broader numerical exploration spanning the historical range of $Z$.

\section{Conclusion} \label{sec_conclusions}

We performed a comprehensive FEM simulation campaign to mechanistically link the deterministic behavior of single localized imperfections to the stochastic response of cylindrical shells containing multiple defects. We established that the Batdorf parameter $Z$ is the governing geometric scaling parameter, collapsing the critical knockdown factor $\kappa$ onto a master curve across the tested radius-to-thickness and height-to-radius ratios. We mapped the $Z$--$\overline{\delta}$ parameter space and identified the transition from immediate global collapse to stable localized buckling. We further showed that $\kappa$ for a single defect is location-dependent, reflecting both edge stiffening and the broad spatial reach of the characteristic butterfly buckling pattern. These single-defect results are the first manifestation of the butterfly defect effect, a localized dimple that buckles into a non-axisymmetric mode whose diagonally extended wings span roughly 8 to 10\% of the shell surface, so that a nominally localized imperfection exerts a non-localized influence on the surrounding structure.

Building on this deterministic baseline, we investigated defect-defect interactions. Pairwise interactions between localized dimples were recently examined by \citet{liu_interaction_2026}, who characterized their dependence on spacing and orientation. We found that the cylindrical shell does not follow a simple weakest-link mechanism. Instead, two defects interact most strongly in an orientation-dependent manner, driven by the constructive interference of their butterfly displacement fields, which merge into a single, more damaging geometric imperfection. This spatial coupling makes relative defect positioning critical to the structural response, and it identifies the butterfly mode as the mechanism underlying the orientation dependence.

Finally, we quantified the statistical size effect governing stochastic distributions of multiple defects. As the mean number of defects $\overline{N}$ increases, the mean knockdown factor $\overline{\kappa}$ decreases asymptotically while the coefficient of variation $\omega_{\kappa}$ decays exponentially. Through the butterfly defect effect, longer cylinders, which host larger defect populations, are more likely to contain a critically interacting cluster, resulting in a lower, more deterministic buckling strength. The same behavior is qualitatively consistent with the reduced scatter observed when historical experimental data are plotted against the dimensionless cylinder length $H/t$.

While this progressive investigation provides a robust mechanistic framework for interpreting the high variance in historical experimental datasets, our numerical model isolated the influence of amplitude variance by using localized dimples of fixed defect size and aspect ratio. Consequently, the knockdown factors observed in this study remain well above the historical floor: a single dimple plateaus near $\kappa \approx 0.53$, and the stochastic ensembles converge to $\kappa \approx 0.55$. Even the most extreme two-defect configuration we examined, a pair of deeply distorted dimples spaced more closely than the ensembles permit, reaches only $\kappa \approx 0.35$, whereas historical experiments report values as low as 0.1 [Fig.~\ref{Fig_Historic_Data}(a)]. A Gaussian dimple closely matches the isolated, axisymmetric buckling state of a spherical shell, making it nearly the worst-case imperfection in that geometry. By contrast, on cylinders, the localized state is the diagonally extended butterfly, so a compact dimple turns out to be a poorer match to the mode the shell forms and may not be its worst-case imperfection. Koiter's general asymptotic theory identifies axisymmetric sinusoidal imperfections as the worst-case imperfection that severely degrades $\kappa$ \cite{koiter_over_1945, gerasimidis_establishing_2018}, though such highly correlated global patterns are unrealistic in actual manufacturing.

In realistic manufacturing, both the depth and the geometric footprint of an imperfection vary from defect to defect, likely presenting as a complex stochastic combination of inward and outward deviations rather than dimples of a single fixed shape. Because larger or elongated imperfections alter edge proximity and reshape the overlap of the butterfly displacement fields that governs their interaction, this variance would likely compound the observed statistical size effect. Nevertheless, the spatial-coupling mechanism and statistical scaling established here remain robust. Identifying the defect classes that set the worst-case knockdown factor for cylinders, and that may approach the low values seen in practice, remains a central open challenge. Extending this framework to richer defect classes, and to the coupled distributions of defect size, shape, and amplitude that describe real structures, is a critical next step toward reliability-based design criteria for imperfection-sensitive cylindrical shells. This extension spans the full progression established here, from the deterministic sensitivity of individual defects, through their pairwise interactions, to the statistics of many.\\ 

\noindent \textbf{Acknowledgments.} We are grateful to Sam Tucker for valuable discussions.\\

\noindent \textbf{Competing interests.} The authors declare no competing interests.\\

\noindent\textbf{Declaration of Generative AI Usage}. During the preparation of this work, the authors used Claude.ai, Gemini.google.com, and Grammarly.com to improve the text's quality and ensure consistency in terminology. After using these tools/services, the authors reviewed and edited the content as needed, and they take full responsibility for the content of the published article.\\

\clearpage
    \appendix
    
\section{Sensitivity to Defect Shape and Size}\label{App_DefectShapeSize}

This appendix evaluates the sensitivity of the knockdown factor $\kappa$ to defect geometry and justifies the standardized profile $(a,\lambda)=(1,1)$ used for the stochastic study in the main text.

\begin{figure}[hbt!]
    \centering
    \includegraphics[width=0.99\linewidth]{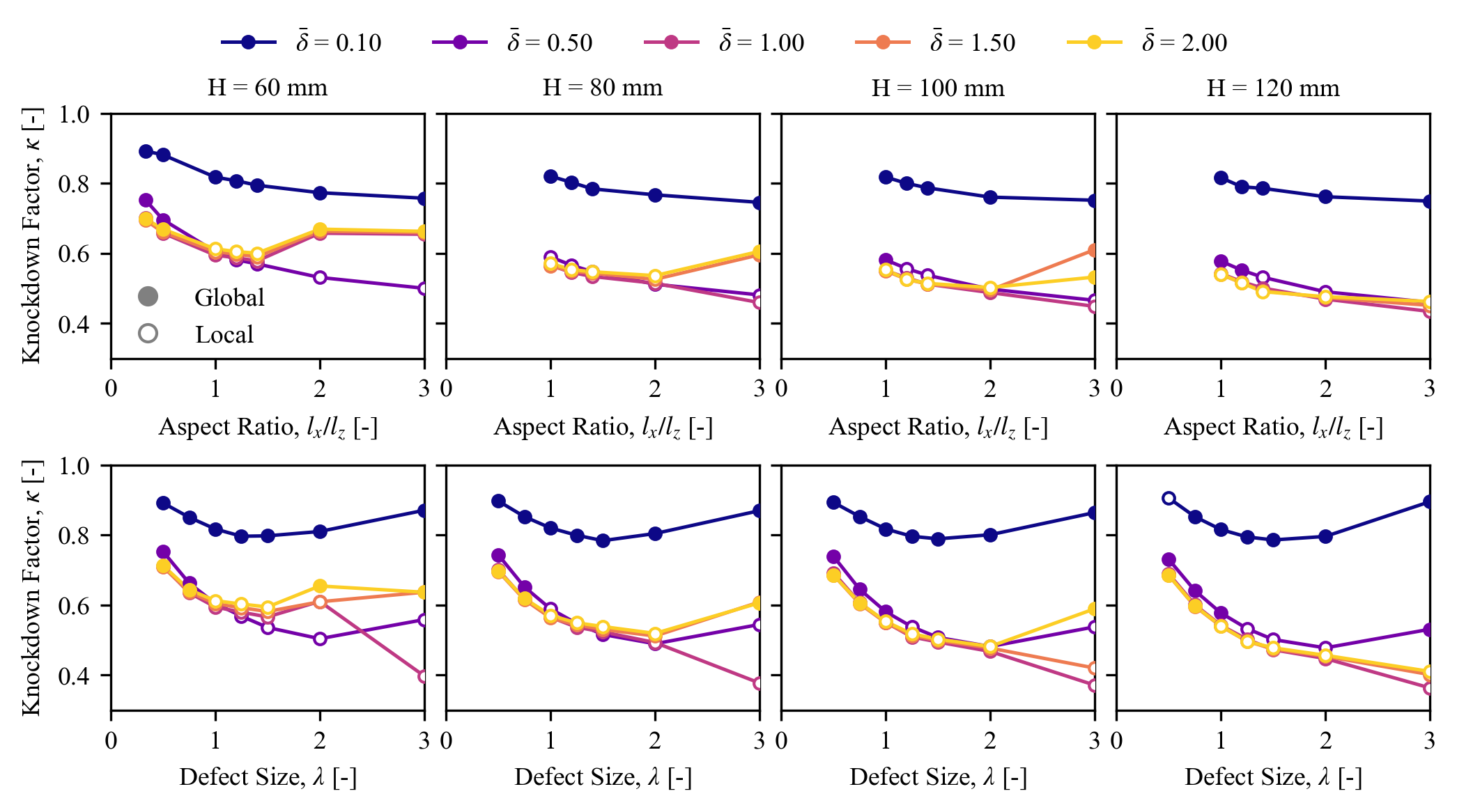}
    \caption{Influence of defect geometry on the knockdown factor ($\kappa$) of cylindrical shells. The top row presents the critical knockdown factor as a function of the defect aspect ratio ($l_x/l_z$), while the bottom row shows the effect of the non-dimensional defect size ($\lambda =l_z/l_c$). Each column corresponds to a specific cylinder height ($H$). The various curves represent different normalized imperfection amplitudes ($\overline{\delta}$).}
    \label{FigA_DefectSize_and_AspectRatio}
\end{figure}

\paragraph{Sensitivity to aspect ratio}
For a fixed defect size $\lambda=1$, the top row of Fig.~\ref{FigA_DefectSize_and_AspectRatio} shows $\kappa$ as a function of the aspect ratio $a=l_x/l_z$. For shallow defects ($\overline{\delta}\le 0.5$), $\kappa$ decreases monotonically toward a plateau as the defect widens circumferentially. For deeper defects ($\overline{\delta}\ge 1$), increasing $a$ beyond a threshold value produces a non-monotonic response in cylinders with $H\lesssim 100\,$mm. This shift reflects a change in buckling mechanism: highly elongated defects produce a gradual loss of pre-buckling stiffness rather than a localized snap-through instability. Taller cylinders delay this transition, sustaining the snap-through mode to higher aspect ratios.

\paragraph{Sensitivity to defect size}
Fixing the aspect ratio at $a=1$, the response to the defect size (plotted as $\lambda = l_z/l_c$) mirrors the aspect-ratio trends (Fig.~\ref{FigA_DefectSize_and_AspectRatio}, bottom row). Shallow defects retain a high knockdown factor that varies only weakly with $\lambda$, whereas deeper defects depend more strongly on it and, for most amplitudes, pass through a minimum near $\lambda \approx 1.5$ to $2$. Beyond this minimum, a further increase in $\lambda$ alters the buckling mode, producing an apparent recovery or scatter in the load capacity. \\

Exceptionally large ($\lambda \gg 1$) or wide ($a \gg 1$) defects introduce mixed buckling regimes and complex post-buckling paths. We therefore restrict the primary multi-defect study to the baseline $(a,\lambda)=(1,1)$, ensuring that the statistical scatter in the stochastic simulations reflects spatial interactions among defects rather than changes in the buckling mechanics of individual defects.

\section{Influence of Defect Size on Edge Interactions}\label{App_BoundarySize}

Here, we evaluate how defect size alters its interaction with the clamped edges. Larger imperfections produce broader buckling displacement fields, so edge interactions appear at greater distances from the edge.

\begin{figure}[h!]
    \centering
    \includegraphics[width=0.85\linewidth]{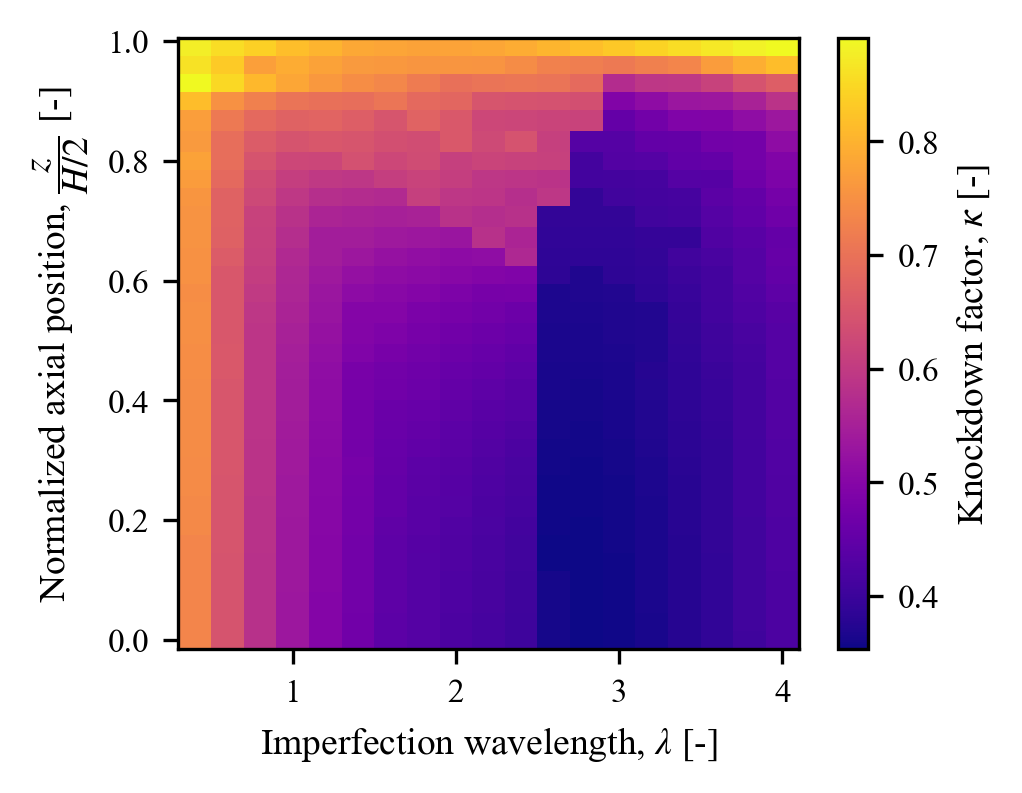}
    \caption{Knockdown factor $\kappa$ for a shell containing a single Gaussian dimple ($\overline{\delta}=1$, $a=1$), mapped across the defect size $\lambda$ and the normalized axial position $z/(H/2)$. The larger geometric footprint of bigger defects pushes the edge-stiffening effect further inward, toward the midplane of the cylinder.}
    \label{Fig_App_BoundarySize}
\end{figure}

For fixed defect parameters $(\overline{\delta}, a) = (1, 1)$, Fig.~\ref{Fig_App_BoundarySize} maps the knockdown factor $\kappa$ against the normalized axial position $z/(H/2)$ and the defect size $\lambda$. For spatially confined defects ($\lambda \lesssim 1.0$), $\kappa$ is independent of axial placement until the defect center reaches $z/(H/2) \approx 0.8$. Edge stiffening is then highly localized: the imperfection must sit almost directly against the edge to artificially raise the buckling load.

As the wavelength increases ($\lambda > 2$), the edge influences a larger portion of the shell. The extended wings of the buckling mode reach the clamped edges prematurely;  for a defect with $\lambda = 3$, the edge raises $\kappa$ already at $z/(H/2) \approx 0.6$. This coupling shows that the shell must provide enough unconstrained surface for the displacement field to develop fully, so that the isolated snap-through instability is captured accurately.

\clearpage 
\section{Interaction of Defects of Unequal Amplitudes}
\label{App_UnequalDefects}

\begin{figure}[h!]
    \centering
    \includegraphics[width=0.95\linewidth]{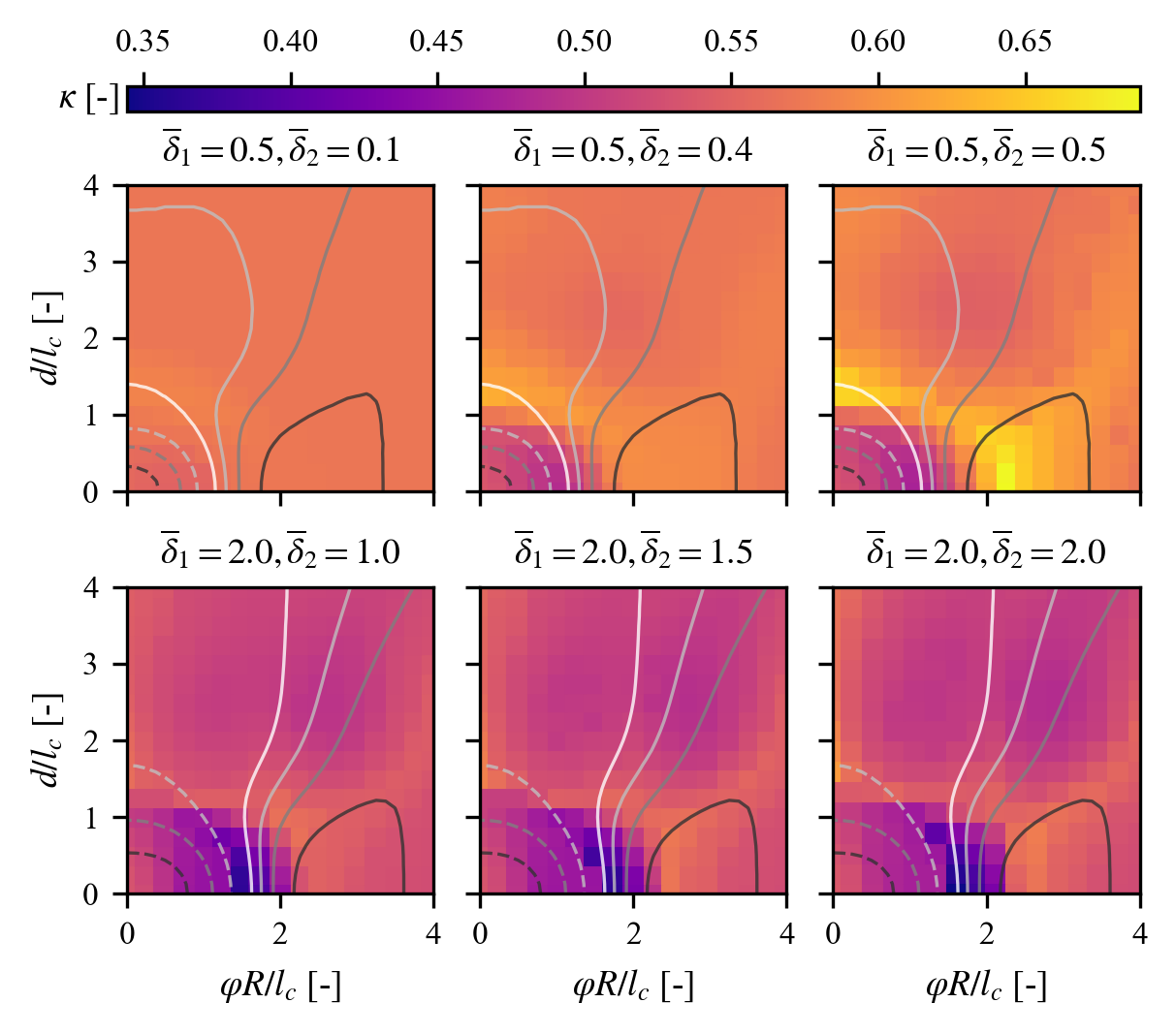}
    \caption{Knockdown factor $\kappa$ for two interacting defects of unequal amplitude, mapped over the normalized circumferential ($\varphi R/l_c$) and axial ($d/l_c$) separation space. The top row shows the pre-plateau regime ($\overline{\delta}_1 = 0.5$) at three secondary amplitudes $\overline{\delta}_2$; the bottom row shows the plateau regime ($\overline{\delta}_1 = 2$). Overlaid solid and dashed lines are the contours of the isolated single-defect radial deflection at buckling onset.}
    \label{Fig_App_UnequalDefects}
\end{figure}

This appendix maps defect-defect interactions for pairs of unequal amplitude, mapping the knockdown factor $\kappa$ over the normalized separation space $(\varphi R/l_c,\, d/l_c)$. We split the analysis by the primary-defect depth $\overline{\delta}_1$: the pre-plateau, imperfection-sensitivity regime ($\overline{\delta}_1 = 0.5$) and the plateau regime ($\overline{\delta}_1 = 2$), shown in the top and bottom rows of Fig.~\ref{Fig_App_UnequalDefects}, respectively.

In the pre-plateau regime ($\overline{\delta}_1 = 0.5$, top row), a shallow secondary defect such as $(\overline{\delta}_1, \overline{\delta}_2) = (0.5, 0.1)$ lets the primary defect dominate, leaving the global buckling load nearly uniform across the separation space. As $\overline{\delta}_2$ approaches the primary amplitude ($\overline{\delta}_2 \to \overline{\delta}_1=0.5$), the interaction intensifies and forms a distinct diagonal zone.

In the plateau regime ($\overline{\delta}_1 = 2$, bottom row), the interaction pattern stays consistent across secondary depths $\overline{\delta}_2 \in \{1.0, 1.5, 2.0\}$; deeper secondary defects only sharpen the interaction slightly.

Across all configurations, the regions of strongest interaction align with the contours of the isolated single-defect radial deflection at buckling onset (lines overlaid in each of the maps in Fig.~\ref{Fig_App_UnequalDefects}), confirming that defect coupling is governed by the overlap of the butterfly displacement fields. Where the diagonal wings overlap, the two displacement fields merge into a single, more damaging imperfection and sharply reduce the load-carrying capacity. Other alignments partially cancel the distortion, leaving localized regions where the deviation is small. Because a comparable, well-aligned secondary defect can drive $\kappa$ below the value set by the primary defect alone, these unequal-amplitude pairs reinforce the finding of Sec.~\ref{sec_DefectDefect} that buckling is not governed by a simple weakest-link mechanism.

\FloatBarrier
   
\bibliographystyle{elsarticle-num-names} 
\bibliography{references}

\end{document}